\documentclass[12pt,a4paper]{article}
\pdfoutput=1
\usepackage[T1]{fontenc}
\usepackage[utf8]{inputenc}

\usepackage{jcappub}
\allowdisplaybreaks
 
\newcommand{\vb}[1]{\ensuremath{\mathbf{#1}}}
\newcommand{\avg}[1]{\ensuremath{\left\langle \,#1\, \right\rangle}}

\def\ie{{\em i.e.}~}

\usepackage[margin=1in]{geometry}
\usepackage{graphicx}
\usepackage{amsfonts}
\usepackage{subfigure}
\usepackage{float}
\usepackage{hyperref}
\usepackage[dvipsnames]{xcolor}
\usepackage{amsmath}
\usepackage[normalem]{ulem}
\usepackage{empheq}
\usepackage{amssymb}
\usepackage{tikz}
\usepackage{tabularx}
\newcommand{\class}{{\sc class}}
\newcommand{\camb}{{\sc camb}}

\usepackage{booktabs}

\newcommand{\eq}[1]{Eq.~(\ref{#1})}
\newcommand{\eqs}[2]{Eqs.~(\ref{#1}, \ref{#2})}
\newcommand{\fig}[1]{Figure~\ref{#1}}

\def\lsim{~\rlap{$<$}{\lower 1.0ex\hbox{$\sim$}}}
\def\bsim{~\rlap{$>$}{\lower 1.0ex\hbox{$\sim$}}}

\def\dd{{\rm d}}

\def\be{\begin{equation}}
\def\ee{\end{equation}}
\def\bea{\begin{eqnarray}}
\def\eea{\end{eqnarray}}
\def\ba{\begin{align}}
\def\ea{\end{align}}
\def\bi{\begin{itemize}}
\def\ei{\end{itemize}}

\newcommand{\bn}{{\mathbf n}}

\def\d{{\rm d}}

\def\vs{\mathrm{\bf s}}

  \def\vs{\vec{ s}} 
\def\d{\rm d}

\def\d{{\rm d}}

\def\vd{{\bf{d}}}
\def\vs{{\bf{s}}}
\def\bx{{\bf{x}}}
\def\bk{{\bf{k}}}
\def\bq{{\bf{q}}}

\def\bv{{\bf{v}}}

\def\ba{{\vec{g}}}

\newcommand{\HH}{\mathcal{H} }

\newcommand{\ndv}{{v_{||}}}

\title{The observed galaxy power spectrum in General Relativity}

\author[a,b,1]{Emanuele~Castorina\note{Author names are listed alphabetically.}}
\author[b,c]{and Enea~Di~Dio}

\affiliation[a]{Dipartimento di Fisica 'Aldo Pontremoli' Università degli Studi di Milano, Milan, Italy}
\affiliation[b]{Theoretical Physics Department, CERN, 1211 Geneva 23, Switzerland}
\affiliation[c]{Center for Theoretical Astrophysics and Cosmology, Institute for Computational Science, University of Zurich, Winterthurerstrasse 190, CH-8057 Zurich, Switzerland}

\emailAdd{emanuele.castorina@unimi.it}
\emailAdd{enea.didio@cern.ch}

\abstract{
Measurements of the clustering of galaxies in Fourier space, and at low wavenumbers, offer a window into the early Universe via the possible presence of scale dependent bias generated by Primordial Non Gaussianites.
On such large scales a Newtonian treatment of density perturbations might not be sufficient to describe the measurements, and a fully relativistic calculation should be employed. The interpretation of the data is thus further complicated by the fact that relativistic effects break statistical homogeneity and isotropy and are potentially divergent in the Infra-Red (IR). 
In this work we compute for the first time the ensemble average of the most used Fourier space estimator in spectroscopic surveys, including all general relativistic (GR) effects, and allowing for an arbitrary choice of angular and radial selection functions. We show that any observable is free of IR sensitivity once all the GR terms, individually divergent, are taken into account, and that this cancellation is a consequence of the presence of the  Weinberg adiabatic mode as a solution to Einstein's equations. We then study the importance of GR effects, including lensing magnification, in the interpretation of the galaxy power spectrum multipoles, finding that they are in general a small, less than ten percent level, correction to the leading redshift space distortions term.  This work represents the baseline for future investigations of the interplay between Primordial Non Gaussianities and GR effects on large scales and in Fourier space.}

\begin{document}

  \begin{minipage}{.45\linewidth}
    \begin{flushleft}
    \end{flushleft}
  \end{minipage}
\begin{minipage}{.45\linewidth}
\begin{flushright}
 {CERN-TH-2021-087}
 \end{flushright}
 \end{minipage}

\maketitle

\section{Introduction}
\label{sec:intro}
The cosmological interpretation of galaxy clustering data is complicated by the fact we observe the angular position and redshift of the galaxies, rather than their true physical position on the lightcone. The mapping between these two systems of coordinates is distorted by a number of observational effects. 

The most important one is Redshift Space Distortions (RSD), \ie the contribution of the galaxies peculiar velocity, projected along the line of sight (LOS), to their observed redshift. RSD select the observer's location as a special place and therefore break the isotropy and homogeneity of $n$th$-$point correlation functions. On the other hand, RSD give access to information about the velocity field we would not be able to retrieve otherwise. 

Other observational effects, arising from the perturbation of the observed redshifts and angles, are  proportional to the gravitational potential or its gradient and are therefore suppressed on sub-horizon scales with respect to the underlying dark matter density field, while becoming important only on very large scales. The relevant dimensionful parameter is the Hubble scale, which separates sub-horizon modes where Netwonian physics applies from the super-horizon scales where the full General Relativity machinery is at work. For this reason these observational features are usually called general relativistic (GR) effects\footnote{ Let us remark that Einstein's equations are not required to derive the relativistic galaxy number counts, see Refs.~\cite{Yoo:2009,Yoo:2010,Challinor:2011bk,Bonvin:2011bg,Jeong:2011as}. However the cancellation of the infrared divergences discussed in this work relies on the existence of the Weinberg adiabatic mode~\cite{Weinberg:2003sw} as a solution of linearized Einstein's field equations.}. Examples of GR effects are lensing, doppler magnification and the Integrated Sachs-Wolfe (ISW) effect.
Galaxy clustering in a relativistic framework has been first derived in Refs.~\cite{Yoo:2009,Yoo:2010,Challinor:2011bk,Bonvin:2011bg,Jeong:2011as} in linear perturbation theory and then extended to higher orders in Refs.~\cite{Yoo:2014sfa,Bertacca:2014dra,DiDio:2014lka,DiDio:2018zmk,DiDio:2020jvo}. It has been also generalized to vector perturbations~\cite{Durrer:2016jzq}, non-flat FLRW universe~\cite{DiDio:2016ykq}, intensity mapping~\cite{Hall:2012wd} and Ly-$\alpha$ forest~\cite{Irsic:2015nla} observables.

GR effects are important for two main reasons. The first one is that they provide a fully gauge-invariant framework to test gravity on the largest scales. 
Secondly, GR effects could be a contaminant to measurements of local Primordial Non Gaussianities (PNG) in the clustering of galaxies with scale dependent bias~\cite{Dalal:2008,Matarrese:2008,Slosar:2008hx,Desjacques:2010jw}. Local PNG appear as large scale divergences, proportional to the primordial gravitational potential,  in the clustering of biased tracers. Local PNG arise from the coupling between the large and small scale modes generated during inflation, and in the squeezed limit they cannot be generated by the dynamics of GR at later times. However, projection effects, like the GR effects responsible for the mapping between true and observed coordinates, \emph{could} mimic PNG, since they also contain terms proportional to the  gravitational potential. The literature on the subject is vast: for photometric surveys measuring angular power spectra, marginalizing over the unknown free parameters that enter the full GR expression causes significant degradation of the error on local PNG~\cite{Camera:2014bwa,Alonso:2015sfa,Fonseca:2015laa,Raccanelli:2015GR}. This is in part due to GR effects contributing mostly to the cross correlation between different redshift bins, which would be zero in a standard Newtonian approach.  In Fourier-space (but in plane-parallel approximation and neglecting integrated effects)Ref.~\cite{Yoo:2012se} concluded that the degeneracy is reduced, especially if one is able to set priors on evolution and magnification biases, due to the different redshift evolution between local PNG and relativistic effects.
A deeper understanding of this issue is therefore of utmost importance for future surveys that could improve over current CMB~\cite{plancknG} and Large-Scale Structure~\cite{Castorina2019} constraints by an order of magnitude~\cite{spherex}.

On this topic, it has been recently argued in~\cite{Grimm:2020ays} that GR effects do not produce any PNG-like feature at large scales in the galaxy power spectrum. This is interpreted as a consequence of the equivalence principle. 
The cancellation of these terms holds for the variance of the galaxy density field, and for a galaxy power spectrum the authors in \cite{Grimm:2020ays} define such that its variance corresponds to the observed one.

The goal of this work is to clarify on the issue of whether GR effects could be degenerate with a measurement of local PNG. 
We will confirm the result in~\cite{Grimm:2020ays} for the variance of the galaxy density field, but at the same time we will show the power spectrum as defined in~\cite{Grimm:2020ays} is not observable, and that true observed power spectrum as estimated in galaxy redshift surveys receives contributions proportional to the gravitational potential even in the absence of PNG.  However, these terms do not show any sensitivity to long wavelength modes, which we will demonstrate is a consequence of the equivalence principle and of the existence of Weinberg adiabatic modes~\cite{Weinberg:2003sw}. Contributions to the power spectrum arising from local PNG are instead genuinely sensitive long wavelength modes. 
For the first time we will also include observational effects like an arbitrary survey geometry and selection function when discussing the importance of GR effects on the power spectrum.
Differently than previous analyses in Fourier space, our approach will include also wide-angle and evolution effects consistently.

This paper is organized as follows. In section~\ref{sec:estimator} we introduce the power spectrum estimator and we discuss its interpretation and its connection to the theoretical modeling. In section~\ref{sec:GR_theory} we describe the fully relativistic galaxy number counts and then in section~\ref{sec:IR_div} we discuss the cancellation of the IR divergences in the correlation function. In section~\ref{sec:GR_effects_Pk} we compute the power spectrum including all the relativistic effects and then we conclude in section~\ref{sec:conclusions}. Several appendices contain the details about azimuthally symmetric window functions (App.~\ref{sec:Azymuthal_window}), the full relativistic correlation function (App.~\ref{sec:correlation_function_appendix}) and the contribution of the observer velocity (App.~\ref{sec:observer_vel_appendix}).

In this work we adopt the following Fourier convention for the Fourier transform
\be
f\left( \bx \right) = \int \frac{\d^3k }{\left( 2 \pi \right)^3} f \left( \bk \right) e^{i \bx \cdot \bk}  \, , \qquad
f\left( \bk \right) = \int \dd^3x f \left( \bx \right) e^{-i \bx \cdot \bk} \, ,
\ee
and the Hankel transform
\be
\xi_\ell \left( s \right) = i^\ell \int \frac{\d k }{2 \pi^2} k^2 P_\ell \left( k \right) j_\ell \left( k s \right)  \, , \qquad
P_\ell \left( k \right) = 4 \pi \left( - i \right)^\ell \int \d s  s^2 \xi_\ell \left( s \right) j_\ell \left( k s \right) \, .
\ee

\section{The power spectrum estimator}
\label{sec:estimator}
Given a catalog  $N(\vb{n},z)$ of galaxy positions in the sky, $\vb{n}$, and redshifts $z$,  we have several ways to construct a quadratic estimator of the data. For the remainder of this work we will focus on spectroscopic surveys, where redshift accuracy is obtained for each object in the catalog. 
If one wants to work in the observed coordinates, then a spherical harmonics approach leads to the measurements of as many angular power spectra as redshifts bins of the data (plus their cross correlation). The advantage of this approach is that the implementation of GR effects is simple and there is no need to assume a fiducial cosmology to perform the measurements, the so called Alcock-Pacinski effect~\cite{AP} \footnote{We have efficient and accurate numerical tools to compute the redshift-dependent angular power spectra including all the relativistic effects: \class{}~\cite{Blas:2011rf,DiDio:2013bqa,Schoneberg:2018fis} and \camb{}source~\cite{Challinor:2011bk}.}. There are two main drawbacks of using angular power spectra. The first one is that, to achieve optimal signal extraction, the number of redshift bins has to be extremely large, see for instance Refs.~\cite{Asorey:2012rd,DiDio:2013sea}. This is more of a concern for small scales analysis than for PNG ones, however in the latter case one has to compute the cross correlation between all the different bins.
This results in a large data vector, which makes the estimation of the covariance matrix of the data a challenging task.  The second problem is that the connection to the perturbative approaches beyond linear theory, required to properly model the small scales, is non trivial\footnote{See~\cite{Castorina2018,Jalilvand19} for a recent solution to this issue.}. 
Similar conclusions hold for spherical-Fourier-Bessel (sFB) methods~\cite{Fisher94,HT95,Tadros99,Percival2004,Rassat:2011aa,Yoo:2013tc,Samushia,Gebhardt21}, where the redshift coordinate is radially Fourier transformed.  

The other main option to build a quadratic estimator of the data is to work in a 3D coordinate system and measure the two-point correlation function or the power spectrum. This approach assumes a fiducial cosmology to convert angular positions and redshifts into distances. This causes no loss of information, but it requires some care if the assumed fiducial cosmology is very different than the true underlying one. That is however not a worry since Planck~\cite{Aghanim:2018eyx} and BAO~\cite{Auburg2015,Alam:2020sor} measurements have constrained the distance-redshift relation to a few percent. The advantages of 3D methods are the straightforward connection to perturbation theory and the reduced dimensionality of data vector, comprising of approximately 100 data points. 
The main drawback of 3D estimators is that GR effects are not easily implemented, especially the ones integrated along the line of sight, like lensing magnification, ISW and time-delay. One of the main goal of this work is to resolve this issue as well, providing the exact interpretation of the estimator of the power spectrum. Due to its practical advantages and robust theoretical interpretation the 3D methods should be preferred in spectroscopic surveys, and Fourier ones in particular to constraints on PNG.

A generic estimator for the multipoles of the power spectrum can be written by summing over all pairs of galaxies with the appropriate Fourier phases,
\be \label{eq:def_estimator}
\hat P_L \left( k \right) = \frac{2 L +1}{ A} \int \frac{d \Omega_\bk}{4 \pi}\int \dd^3 s_1  \dd^3s_2 \, \Delta \left(\vs_1 \right) \Delta \left( \vs_2 \right) \phi \left( \vs_1 \right) \phi \left( \vs_2 \right) e^{i \bk \cdot \left( \vs_1 - \vs_2 \right) }
\mathcal{L}_L \left( \hat \bk \cdot \hat \vd_{\rm LOS} \right)\,,
\ee
where $A$ is a normalisation constant, $\Delta(\vs)$  denotes the galaxy density contrast, $\phi \left( \vs \right)$ is the survey window function, and $\mathcal{L}_L$ is the Legendre polynomial of order $L$.
This is the so called Yamamoto estimator~\cite{Yamamoto:2005dz}, which reduces to the FKP  estimator~\cite{FKP} for $L=0$. 
Beyond the monopole, the estimator depends explicitly on the choice of the line of sight, $ \vd_{\rm LOS}$, for the pair of galaxies at $\vb{s}_1$ and $\vb{s}_2$.

For an unbiased estimate of cosmological parameters it is therefore of utmost importance to first understand the relation between the average of the estimator and the analytical prediction. 
Some of the following considerations have appeared elsewhere before \cite{Castorina:2017inr,Scoccimarro2015}, but we report them here for clarity. In Section~\ref{sec:intro} we said that RSD, and other GR effects, break homogeneity and isotropy of correlation functions, making for instance the power spectrum of the observed galaxy density field non-diagonal. This means that if we Fourier Transform $\Delta[\vs(\vb{n},z)]$ to $\Delta(\vb{k})$, we obtain
\begin{align}
 \avg{\Delta(\vb{k}_1) \Delta(\vb{k}_2) } \equiv P(\vb{k}_1,\vb{k}_2)\,,
\end{align}
while the Yamamoto estimator defined above is a function of a single wavenumber $k$. We will now show that the relation of $P(\vb{k}_1,\vb{k}_2)$ with $\hat{P}_L \left( k \right)$ is LOS-dependent. Consider first the midpoint as the line of sight, $\vd_{\rm LOS}= \vd_{\rm m} \equiv (\vs_1+\vs_2)/2$, and assume that the window function is equal to unity everywhere inside the survey volume, which we take to cover the whole sky. It is easy to see that in this case the average of the estimator is given by~\cite{Scoccimarro2015,Castorina:2017inr}
\begin{align}
\avg{\hat{P}_L(k)} = & \frac{(2 L+1)}{A}\int\mathrm{d}^3d_{\rm m}\,\int \frac{\mathrm{d}^3 q}{(2\pi)^3} e^{i\vb{q}\cdot\vb{d}_{\rm m}} \notag \\
&\times\int \frac{\mathrm{d}\Omega_{\vb{k}}}{4\pi} P(\vb{k}+\vb{q}/2,-\vb{k}+\vb{q}/2)\mathcal{L}_L(\hat{\vb{k}}\cdot \hat{\vb{d}}_{\rm m})\,,
\end{align}
revealing that the Yamamoto estimator further compresses the underlying power spectrum by averaging over the wavenumber $\vb{q}$ associated to the line of sight $\vb{d}_m$. Another possible choice for the line of sight is one of the two pair members, $\vd_{\rm LOS} = \vs_1$, usually called the endpoint line of sight. In this case 
\be
\avg{ \hat P_L \left( k \right) } =
 \frac{2 L +1}{ A} \int \frac{\dd \Omega_{\mathbf{k}}}{4 \pi}\int \dd^3 s_1  \int \frac{\dd^3 q}{\left( 2 \pi \right)^3} \, P \left( \bq - \bk , \bk \right)  e^{i \bq \cdot \vs_1  }
\mathcal{L}_L \left( \hat \bk \cdot \hat \vs_1 \right),
\ee
which shows how different choices of LOS result in different weighting of the momentum~$\mathbf{q}$.

A more compact expression for the average of the estimator can be obtained by noticing that the configuration space two-point correlation function depends on the triangle formed by the observer and the pair of galaxies
\begin{align}
 \avg{\Delta(\vb{s}_1) \Delta(\vb{s}_2) } \equiv \xi(\vb{s}_2-\vb{s}_1,\vb{d}_{\rm LOS})\,
\end{align}
and it can therefore be formally Fourier transformed with respect to $\vb{s} \equiv \vb{s}_2-\vb{s}_1$ to define a LOS dependent power spectrum
\begin{align}
P(\vb{k},\vb{d}_{\rm LOS}) \equiv \int \dd ^3 s\,e^{-i \vb{k}\cdot\vb{s}} \xi(\vb{s},\vb{d}_{\rm LOS})\,.
\end{align}
The Yamamoto estimator is then the LOS average of the LOS dependent power spectrum 
\begin{align}
 \avg{\hat{P}_L(k)} =   \frac{(2L+1)}{A} \int \frac{\mathrm{d}\Omega_\vb{k}}{4\pi}\, \int \mathrm{d}^3 d_{\rm LOS} \, P(\vb{k},\vb{d}_{\rm LOS}) \mathcal{L}_L(\hat{\vb{k}}\cdot\hat{\vb{d}}_{\rm LOS}) \, .
\end{align}
The above expression also clarifies how to take the plane-parallel limit of the estimator~\cite{Castorina:2017inr}. 

\subsection{The convolution with a window function}
For practical reasons the choice $\vd_{\rm LOS} = \vs_1$ is usually preferred~\cite{Bianchi2015,Scoccimarro2015,Hand2017}, since it requires fewer Fourier Transforms than the identification of the LOS as the bisector between the two galaxies and the observer~\cite{Castorina:2017inr}. 
We shall use this choice, usually referred as the endpoint LOS in the literature, for the remainder of this work.
Although not strictly necessary, in our implementation we will assume that the window function can be separated into a radial and angular part, \ie $\phi(\vb{s}) = \phi(s)W(\hat{\vb{s}})$. Generalizations to non-separable masks are straightforward.
If we start from the following decomposition for the two-point correlation function
\begin{align}
    \xi(\vb{s},\vb{s}_1) = \sum_\ell \xi_\ell(s,s_1) \cal{L}_\ell(\mu)\,,
\end{align}
with $\mu \equiv \hat{\vb{s}}\cdot\hat{\vb{s}}_1$, we arrive to a compact expression for the expectation value of the Yamamoto estimator
\bea
\label{eq:estimator_standard}
\langle \hat P_L \left( k \right) \rangle &=& \frac{2 L +1}{ A}\sum_\ell \left( -i\right) ^L \int \dd^3 s_1  \dd^3s    \xi_\ell \left( s, s_1  \right) \mathcal{L}_\ell \left( \hat \vs_1 \cdot \hat \vs  \right)  
\mathcal{L}_L \left( \hat \vs \cdot \hat \vs_1 \right) j_L \left( k s \right) \phi\left( \vs_1 \right) \phi\left( \vs_2 \right) 
\nonumber \\
&=&
\frac{2 L +1}{A }\sum_{\ell \ell_1} \left( -i\right) ^L   
\left(
\begin{array}{ccc}
L & \ell & \ell_1 \\
0 &0&0
\end{array}
\right)^2 \left( 2 \ell_1 +1 \right) 
\nonumber \\
&& \qquad
 \int \dd^3 s_1 \dd^3s    \xi_\ell \left( s, s_1  \right)  
\mathcal{L}_{\ell_1} \left( \hat \vs \cdot \hat \vs_1 \right) j_L \left( k s \right) \phi\left( \vs_1 \right) \phi\left( \vs_2 \right) 
\nonumber \\
&=&
\frac{2 L +1}{A }\sum_{\ell, \ell_1} \left( -i\right) ^L
\left(
\begin{array}{ccc}
L & \ell & \ell_1 \\
0 &0&0
\end{array}
\right)^2 \left( 2 \ell_1 +1 \right) 
\nonumber \\
&& \qquad
 \int \dd s_1 s_1^2 \dd s s^2  \xi_\ell \left( s, s_1  \right)  j_L \left( k s \right)  \phi\left( s_1 \right) F_{\ell_1}\left( s_1 , s \right) 
\, ,
\eea
where we have introduced 
\be
\label{eq:Fell_def}
F_{\ell_1} \left( s_1 , s \right) = \int \dd\Omega_{\hat \vs} \dd\Omega_{\hat \vs_1} \phi \left( s_2 \right) W \left( \hat \vs_1 \right) W \left( \hat \vs_2 \right)  \mathcal{L}_{\ell_1} \left( \hat \vs_1 \cdot \hat \vs  \right)   \, .
\ee 
The integration variable $s_1$ serves to keep track of the redshift evolution of the sample within the survey, and of the different dependence of the GR effects on the distance between the observer and the galaxies. In real surveys the functions $F_\ell(s_1,s)$ can be estimated using traditional pair counting algorithms or Fast Fourier Transforms (FFT). 
\subsubsection{The effective redshift}
A useful approximation often employed in the literature is to assume that the theoretical prediction can be computed at some effective redshift. This allows to take the correlation function multipoles outside of the $\dd s_1$ integral in Eq.~(\ref{eq:estimator_standard}), that can be computed once and for all over the window function multipoles $F_\ell$. 
The effective redshift approximation is motivated by the fact that we are mostly interested in the clustering at separations much smaller than the comoving distance associated to the redshift of the individual galaxies, that can therefore be modeled at the same redshift. 
On the small scales relevant for BAO and RSD this has been shown to be an excellent approximation~\cite{Vlah2016}, as well for PNG studies on small patches of the sky~\cite{Castorina2019}. 
To see how this works let us assume that the correlation function can be written as a sum over different terms, each of them being the correlation function between two operators $\cal{O}$ and $\cal{O}'$ entering the expansion of the galaxy density field,
\begin{align}
\xi_\ell(s,s_1) \equiv \sum_{\cal{O}\cal{O}'}\xi_{\ell,\cal{O}\cal{O}'}(s,s_1)\,.
\end{align}
For each contribution to the correlation function we could therefore choose an effective redshift such that at a given, small enough, reference separation $\tilde{s}$, the model evaluated at this suitably defined redshift exactly matches the full prediction in Eq.~(\ref{eq:estimator_standard}). In practice, at small scales the correlation function is dominated by the Newtonian terms in the plane parallel limit, that allows us to define a single effective redshift for each multipole 
\begin{align} \label{eq:def_zeff_th}
  \xi_{\ell,}(\tilde{s},z_{\rm eff,\ell})\equiv  \frac{  \int \dd^3 s_1  \,\xi_{\ell}(\tilde{s},s_1) \phi(s_1)^2 W(\hat{s}_1)^2 }{\int \dd ^3s_1 \,   \phi(s_1)^2 W(\hat{s}_1)^2} \,.
\end{align}
Notice this operation is always possible as long as the transfer function and the growth factors are separable.  
Under the effective redshift approximation the average of the Yamamoto estimator reads 
\begin{align}
    \avg{\hat{P}_L(k)} = &\frac{2 L +1}{A }\sum_{\ell, \ell_1} \left( -i\right) ^L
\left(
\begin{array}{ccc}
L & \ell & \ell_1 \\
0 &0&0
\end{array}
\right)^2 \left( 2 \ell_1 +1 \right)   \int  \dd s s^2  \,\xi_\ell \left(s  , z_{\rm eff,\ell}\right )  j_L \left( k s \right) \nonumber  \\& \times \int \dd s_1 s_1^2 
\, \phi\left( s_1 \right) F_{\ell_1}\left( s_1 , s \right) \nonumber \\
    \equiv & \frac{2 L +1}{A }\sum_{\ell, \ell_1} \left( -i\right) ^L
\left(
\begin{array}{ccc}
L & \ell & \ell_1 \\
0 &0&0
\end{array}
\right)^2 \int  \dd s\, s^2  \xi_\ell \left(s , z_{\rm eff,\ell} \right)  j_L \left( k s \right) Q_{\ell_1}(s) \, .
\label{eq:zeff_theory}
\end{align}
The functions $Q_{\ell_1}$ can be easily estimated using FFT methods~\cite{Beutler:2018vpe}.
It is often the case that the different multipoles have very similar $z_{\rm eff,\ell}$, and therefore a common effective redshift is defined for the entire correlation function\footnote{Another possible definition is the pair weighted effective redshifts, see for example \cite{deMattia:2020fkb}.},
\begin{align}\label{eq:def_zeff}
z_{\rm eff} = \frac{\int\dd^3 s\, \phi(s)^2 W(\hat{s})^2 z(s)}{\int \dd^3 s\, \phi(s)^2 W(\hat{s})^2}\,.
\end{align}
The definition above does not guarantee that the small scale power spectrum evaluated at $z_{\rm eff}$ converges to the true answer, but the difference can usually be reabsorbed by changing the value of the unknown bias parameters. A global definition is also more prone to anisotropies in the orientation of galaxy pairs in the survey, which could also bias the theoretical prediction, as recently pointed out in~\cite{Obuljen:2021ryv}.

The simple picture described above is complicated by those correlation function terms which are integrated along the LOS, like lensing magnification and ISW. In this case, the definition of effective redshift given above does not strictly hold. Nevertheless we will see in the next sections (see Figs.~\ref{fig:eff_redshift} and~\ref{fig:GR_zeff_effects_ratio}) that our choices in \eqs{eq:def_zeff_th}{eq:def_zeff} are accurate enough for these terms as well, vastly simplifying the convolution with the window function.  

\subsubsection{The Integral Constraint}
In redshift surveys the mean number density of the underlying galaxy population is usually not known a priori, and it is estimated from the data themselves. The resulting overdensity will therefore vanish when integrated over the entire volume of the survey, an effect known as the Integral Constraint (IC). In addition, if the redshift selection function is also estimated from the data, the overdensity has to vanish in radial bins when averaging over the angular footprint. This is the so called Radial Integral Constraint (RIC). The mathematics behind the two effects is very similar, therefore we will concentrate only the global IC. 
We follow Ref.~\cite{deMattia:2019vdg}, who showed that, to a very good accuracy, imposing the global IC boils down to following replacement of the theory prediction 
\be
\avg{\hat{P}_L \left( k \right)} \rightarrow \avg{\hat{P}_L \left( k \right)} - \avg{\hat{P}_0 \left( 0 \right)} \frac{Q_L\left( k \right)}{Q_0 \left( 0 \right) }  \equiv \avg{\hat{P}_L \left( k \right)}- P_L^{\rm ic} \left( k \right) \, ,
\ee
where 
\be
Q_L\left( k \right) = 4\pi \left( -i \right)^L \int \dd s \,s^2 \, j_L(ks) Q_L(s)
\ee
Unless otherwise noted, all the Figures and comparisons shown in the next Sections always include the global IC in the theoretical prediction.

%%%%%%%%%%%%%%%%%%%%%%%%%%%%

\section{The relativistic galaxy number density}
\label{sec:GR_theory}
In the previous section we laid out the formalism required to compute the measured power spectrum. The only missing ingredient is a model for the correlation function $\xi(s,s_1,\mu)$ including all the relativistic effects.
From the galaxy catalog $N \left( \bn , z \right)$ in terms of the observed angle $\bn$ and redshift $z$, we can defined the so-called galaxy number counts
\be
\Delta \left( \bn , z \right) = \frac{N \left(\bn , z \right) - \langle N \left( \bn , z \right) \rangle}{\langle N \left( \bn , z \right) \rangle}
\ee
where $\langle .. \rangle$ denotes the angular average at fixed observed redshift. Being $\Delta \left( \bn , z \right)$ an observable quantities we can express it in any gauge. We adopt therefore the Newtonian gauge
\be
ds^2 = a^2(\tau) \left[ - \left( 1+ 2 \Psi \right)\dd\tau^2 + (1- 2 \Phi) \dd\vb{x}^2 \right]
\ee
where $\tau$ denotes the conformal time and the scalar metric perturbations $\Psi$ and $\Phi$ are the Bardeen potentials.

The full relativistic number counts to linear order in perturbation theory~\cite{Yoo:2009,Yoo:2010,Bonvin:2011bg,Challinor:2011bk,Jeong:2011as}, including the observer terms~\cite{Scaccabarozzi:2018vux,Grimm:2020ays}, reads\footnote{In Refs.~\cite{Scaccabarozzi:2018vux,Grimm:2020ays} the Authors implicitly assume that the surveys are limited in volume. However, considering that current and upcoming surveys will be limited in flux we need to introduce also the magnification bias $s_b$, defined as the slope of the luminosity function at the luminosity threshold, following the same convention of Refs.~\cite{Challinor:2011bk,DiDio:2013bqa,DiDio:2016ykq}. Therefore by Taylor expanding around the threshold luminosity and considering that the fractional fluctuation of the luminosity is twice the fractional fluctuation of the luminosity distance ($\delta L / \bar L = 2 \delta D_L / \bar D_L$), we need to replace
$$
\Delta \left( \bn , z \right) \rightarrow \Delta \left( \bn , z \right) - 5 s_b\left( z \right) \frac{\delta D_L }{\bar D_L} \, .
$$
To obtain \eq{eq:Delta_rel}, we have used the luminosity distance of Ref.~\cite{Scaccabarozzi:2018vux} to properly include also the terms evaluated at the observer position.
}
\bea
\label{eq:Delta_rel}
\Delta \left( \bn ,z \right) &=& b_1 D_m + \HH^{-1} \partial_r  \ndv   + \frac{5s_b-2}{2} \int_0^{r} dr' \frac{r - r'}{r r'} \Delta_\Omega \left( \Psi + \Phi \right) 
\nonumber \\
&&+ \left( 5 s_b +\frac{2 - 5s_b}{\HH r} + \frac{\dot \HH}{\HH^2} - f_{\rm evo} \right)
\nonumber \\
&& \quad\left(\HH_0 V_o + \Psi - \Psi_o + \ndv - \ndv_o + \int_\tau^{\tau_o} \left( \dot \Psi + \dot \Phi \right) d\tau' \right)
\nonumber \\
&&
+ \left( 5 s_b -2 \right) \Phi + \Psi + \dot \Phi \HH^{-1} + \left( f_{\rm evo} - 3 \right) \HH V { - \frac{2 - 5s_b}{r } V_o  - \left( 2 - 5s_b\right) \ndv_o }
\nonumber \\
&&
+ \frac{ 2 - 5 s_b }{r} \int_\tau^{\tau_o} \left( \Psi + \Phi \right) d\tau' 
\, ,
\eea
where we have only assumed the validity of the Euler equation for baryons and dark matter. In the equation above $V$ denotes the velocity potential and $\ndv= \bn \cdot \bv$, where $\bn$ is the unit vector pointing from the observer to the source, and $\bv$ is the peculiar velocity in Newtonian gauge. We denote the partial derivative with respect to conformal time $\tau$ with a dot. The gauge-invariant density contrast $D_m$ coincides with the density fluctuation in the comoving gauge.  To relation between dark matter and galaxies is parametrized by a galaxy bias $b_1$, a magnification bias $s_b$ and a evolution bias $f_{\rm evo}$.  Terms with a subscript `$o$' indicate perturbations evaluated at the observer position which are needed for the gauge invariance of all the expressions. For example the term $\Psi-\Psi_o$ in the third line of the equation above is the standard gravitational redshift which is proportional to the difference between the gravitational potential at the source and the observer. 

\begin{figure}
\begin{center}
\includegraphics[width=0.5\textwidth]{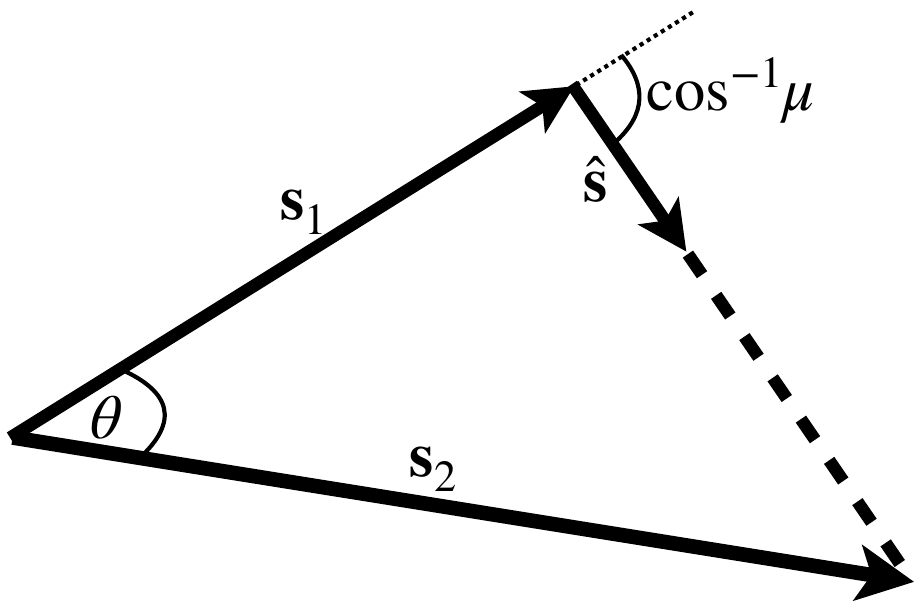}
\caption{We show the systems of coordinates adopted in this work: $(s_1, s_2,\theta)$ and $(s, s_1 , \mu)$. The line of sight variables $\chi_1$ and $\chi_2$ run from the observer to the sources along the vectors $\vs_1$ and $\vs_2$, respectively. \label{fig:draw}}
\end{center}
\end{figure}

We group the different relativistic effects in the following equation as follows: standard density plus RSD (first line), lensing (second line), Doppler (third line), local gravitational potentials (fourth line), integrated gravitational potentials (fifth line),  
\bea
\Delta \left( \bn ,z \right) &=& b_1 D_m + \HH^{-1} \partial_r  \ndv   
\nonumber \\
&&
+ \frac{5s_b-2}{2} \int_0^{r} dr' \frac{r - r'}{r r'} \Delta_\Omega \left( \Psi + \Phi \right) 
\nonumber \\
&&+ \mathcal{R} \left( \ndv - \ndv_o \right) { - \left( 2 - 5s_b\right) \ndv_o }
\nonumber \\
&&+ \bigg\{ \left( \mathcal{R} { - \frac{2-5s_b}{\HH_0 r}}\right) \HH_0 V_o + \left( \mathcal{R} +1 \right)\Psi -\mathcal{R} \Psi_o  
+ \left( 5 s_b -2 \right) \Phi  + \dot \Phi \HH^{-1} 
\nonumber \\
&& \qquad 
+ \left( f_{\rm evo} - 3 \right) \HH V \bigg\}
\nonumber \\
&&
+ \frac{ 2 - 5 s_b }{ r} \int_\tau^{\tau_o} \left( \Psi + \Phi \right) d\tau'
+ \mathcal{R}\int_\tau^{\tau_o} \left( \dot \Psi + \dot \Phi \right) d\tau' \, ,
\eea
where we have introduced the redshift dependent parameter
\be
\mathcal{R} = 5 s_b +\frac{2 - 5s_b}{\HH r} + \frac{\dot \HH} {\HH^2} - f_{\rm evo} \, .
\ee
We can indeed think of the relativistic number counts as a sum of different operators $\mathcal{O}(\bn,z)$, such that the total correlation function requires the computation of all possible terms of the form
\be
\langle \mathcal{O} \left( \bn_1 , z_1 \right) \mathcal{O}' \left( \bn_2 , z_2 \right) \rangle = \xi_{\mathcal{O}\mathcal{O}'} \left( s_1 , s_2, \hat \vs_1 \cdot \hat \vs_1 \right) = \xi_{\mathcal{O}\mathcal{O}'} \left( s, s_1 , \mu \right) \, .
\ee
The correlation function can be computed directly as a function of $(s,s_1,\mu)$, or rotated into this basis from another parametrization, for instance in terms of $(s_1,s_2,\hat{\vb{s}}_1\cdot\hat{\vb{s}}_2)$. A pictorial representation of both coordinates system is shown in Fig.~\ref{fig:draw} We have checked that both methods give identical results, and we will use them interchangeably according to our convenience. For comparison with previous work we will more often work in the  $(s_1,s_2,\hat{\vb{s}}_1\cdot\hat{\vb{s}}_2)$ basis. The latter has the advantage of retaining a simple form even  upon dropping the effective redshift approximation.

So far we have not assumed any theory of gravity. Now, for the rest of the manuscript we will adopt General Relativity. Therefore, by using Einstein equations to relate metric and velocity perturbations to the density fluctuation we can write the correlation function as a linear combination of the functions
\be
\label{eq:Iln}
I^n_\ell \left( s \right) = \int \frac{\dd q }{2 \pi^2} q^2 P \left( q \right) \frac{j_\ell \left( q s \right) }{\left( q s \right)^n} \, ,
\ee
where $P(q)$ is the linear matter power spectrum at $z=0$.
In our notation, we introduce a differential operator $\mathcal{D}_\mathcal{O}$ associated to each $\mathcal{O}$, which allows us to arrive to 
\bea
\label{eq:correlation_differential}
&& \hspace{-1.5cm}\xi_{\mathcal{O}\mathcal{O}'} \left( s_1 , s_2 , \hat \vs_1  \cdot \hat \vs_2 \right) 
\nonumber \\
&=& \int \frac{\dd^3 q}{\left( 2 \pi \right)^3} P\left( q \right) \int_0^{s_1} \dd\chi_1 \int_0^{s_2} \dd\chi_2  \mathcal{D}_\mathcal{O}\left( q, s_1 , \chi_1 \right)  \mathcal{D}_\mathcal{O}'\left( q,s_2 ,\chi_2 \right) e^{i \bq \cdot \left(  \chi_2 \hat \vs_2 - \chi_1 \hat \vs_1\right)}
\nonumber \\
&=&
\int \frac{\dd q}{2 \pi^2 }  q^2 P\left( q \right)\int_0^{s_1} \dd\chi_1 \int_0^{s_2} \dd\chi_2  \mathcal{D}_\mathcal{O}\left( q ,s_1, \chi_1\right)  \mathcal{D}_\mathcal{O}'\left( q,s_2, \chi_2 \right) j_0 \left( q \chi \right) \, ,
\eea
where $\chi= \sqrt{\chi_1^2 + \chi_2^2 - 2 \chi_1 \chi_2 \cos\theta} $. These differential operators, acting on $j_0 \left( q \chi \right) $ will lead to the functions $I^n_\ell$.
To avoid confusion, in this work we will always call $q$ the momentum that enters the calculation of the correlation function, and $k$ the wavenumber associated to the actual measurement of $\hat{P}_L(k)$.
For the different perturbations we have
\bea \label{eq:mapping_start}
\begin{array}{cc}
D_m \left( s_1 \right) \rightarrow  \mathcal{D}_\delta =
T_\delta \left( s_1 \right) \delta_D \left( s_1 - \chi_1 \right)\, , &
\\
%%%%%%%%%%%%%%%%%%%%%%%%%%
\ndv \left( s_1 \right) \rightarrow \mathcal{D}_\ndv= \delta_D \left( s_1 - \chi_1 \right) \frac{T_V \left( s_1 \right)  }{q^2 } \partial_{\chi_1}\, ,
% \\ 
\hspace{0.5cm} & \hspace{0.5cm} 
%%%%%%%%%%%%%%%%%%%%%%%%%%
\ndv_o \rightarrow  \mathcal{D}_{\ndv_o}=
\delta_D \left(  \chi_1 \right)
\frac{T_V \left(0 \right) }{q^2 }  \partial_{\chi_1}  \, ,
\\
%%%%%%%%%%%%%%%%%%%%%%%%%%
V \left( s_1 \right) \rightarrow  \mathcal{D}_V
= \delta_D \left( s_1 - \chi_1 \right)
\frac{T_V \left( s_1 \right) }{q^2} \, , 
% \\ 
\hspace{0.5cm} & \hspace{0.5cm} 
%%%%%%%%%%%%%%%%%%%%%%%%%%
V_o \rightarrow \mathcal{D}_{V_o}
=\delta_D \left(  \chi_1 \right) \frac{T_V \left( 0 \right)}{q^2} \, ,
\\
%%%%%%%%%%%%%%%%%%%%%%%%%%
\Phi \left( s_1 \right)  
\rightarrow  \mathcal{D}_\Phi
= \delta_D \left( s_1 - \chi_1 \right)\frac{T_\Phi \left( s_1 \right) }{q^2}\, , &
\\
%%%%%%%%%%%%%%%%%%%%%%%%%%
\Psi\left( s_1 \right)  \rightarrow  \mathcal{D}_\Psi
= \delta_D \left( s_1 - \chi_1 \right)\frac{T_\Psi \left( s_1 \right) }{q^2}\, , 
% \\ 
\hspace{0.5cm} & \hspace{0.5cm} 
%%%%%%%%%%%%%%%%%%%%%%%%%%
\Psi_o  \rightarrow \mathcal{D}_{\Psi_o} 
= \delta_D \left(  \chi_1 \right)\frac{T_\psi \left( 0\right) }{q^2}\, ,
\\
%%%%%%%%%%%%%%%%%%%%%%%%%%
\Psi\left( \chi_1 \right)  \rightarrow  \mathcal{D}_{ \Psi} =
 \frac{T_{ \Psi} \left( \chi_1 \right)}{q^2}\, ,
\hspace{0.5cm} & \hspace{0.5cm} 
 \dot \Psi\left( \chi_1 \right)  \rightarrow  \mathcal{D}_{\dot \Psi} =
 \frac{T_{\dot \Psi} \left( \chi_1 \right)}{q^2}\, , 
 \\
%%%%%%%%%%%%%%%%%%%%%%%%%%
  \Phi \left( \chi_1 \right)  \rightarrow  \mathcal{D}_{ \Phi} 
=  \frac{T_{\Phi} \left( \chi_1 \right)}{q^2}\,
\hspace{0.5cm} & \hspace{0.5cm} 
 \dot \Phi \left( \chi_1 \right)  \rightarrow  \mathcal{D}_{\dot \Phi} 
=  \frac{T_{\dot \Phi} \left( \chi_1 \right)}{q^2}\, . 
\end{array}
\eea
Einstein gravity will constrain these transfer functions and, in particular in the absence of anisotropic stress, we have a single scalar degree of freedom. In this case the transfer functions are given by
\bea
T_\delta  &=& b_1 D_1 \, , \\
T_V &=& - \HH f D_1 \, , \\
T_\Psi = T_\Phi &=& -\frac{3}{2}\frac{\HH_0^2}{a} \Omega_{M,0} D_1 \, , \\
T_{\dot \Psi}= T_{\dot \Phi} &=&-\frac{3}{2}\frac{\HH_0^2}{a} \Omega_{M,0} D_1 \HH \left( f - 1 \right) \, .
\eea
where $D_1$ is the linear growth factor and $f$ is the linear growth rate.

The galaxy correlation function can be computed for instance using the public code COFFE~\cite{Tansella:2018sld}, which we expanded to include all possible correlations between the observer's and source terms, between perturbations evaluated at the observer location, and to allow the multipole expansion in terms of the endpoint LOS.

The contributions arising from auto correlation of the observer terms represent the variance of these operators as they would be measured by observers with the same $\tau_o$, \ie they account for the difference between a randomly chosen and a preferred observer~\cite{Mitsou:2019ocs}. 
In the cross correlations between the observer and the source what matters are instead the fluctuations on very large scales, comparable to the distance to the source. These long wavelength modes are stochastic in nature and cannot be known a priori. For example in a determination of the observer's velocity, what it is effectively being measured are the small scale fluctuations on the scale of our Local Group, which are by far the largest contribution to the observer's velocity.

\section{Infrared divergences of the correlation function}
\label{sec:IR_div}

\subsection{The variance of the galaxy density field and the $1/q^4$ terms}
If the expression for the galaxy number counts in \eq{eq:Delta_rel} describes a gauge invariant and observable quantity, then all the summary statistics we compute out of it must be finite. 
However, a technical problem is immediately encountered in the calculation of the galaxy correlation functions. Namely, the contribution from any two powers of the metric potentials, for example 
\be
\avg{ \Phi(\vs_1)\Phi(\vs_2)}  =\int_0^\infty \frac{\dd q q^2}{2 \pi^2} \mathcal{D}_\Phi(q,s_1) \mathcal{D}_\Phi(q,s_2) P(q) j_0(q s) \propto  \int_0^\infty \frac{\dd q q^2}{2 \pi^2} q^{-4} P(q) j_0(q s)\,,
\ee
diverges in the Infra-Red (IR) for typical $\Lambda$CDM power spectra, \ie when the lower integration limit is sent to zero.  A healthy theory cannot have such a feature, which could resolve by itself within the framework, or could indicate a breakdown of one of the model assumptions. 

A practical solution would be to remove by hand these divergences. This approach has been considered in~\cite{Tansella:2018sld}, and we will now show that it is intimately related to the exact cancellation of the divergent terms. 
Consider the different operators $\cal{O}$ and $\cal{O}'$ contributing to the divergent part of the correlation function
\begin{align}
    \xi^{\rm div}(s,s_1,\mu) \equiv \sum_{\cal{O}\cal{O}'} \xi^{\rm div}_{\cal{O}\cal{O}'}(s,s_1,\mu)
\end{align}
where each term is proportional to $I^4_0$ defined in \eq{eq:Iln}\footnote{The different terms are proportional to $I^4_0$ for different arguments: $s$, $s_1$, $s_2$, $\chi_1$, $\chi_2$, $\sqrt{s_1^2+ \chi_2^2 - 2s_1 \chi_2 \cos\theta}$, $\sqrt{\chi_1^2+ s_2^2 - 2 \chi_1 s_2 \cos\theta}$ and $\sqrt{\chi_1^2+ \chi_2^2 - 2 \chi_1 \chi_2 \cos\theta}$}.
One way to eliminate the IR divergency is to subtract unity from the spherical Bessel function in the definition of $I^4_0$
\be
\label{eq:Itilde}
I^4_0 \left( s \right) \rightarrow \tilde I^4_0 \left( s \right) = \int \frac{\dd q }{2\pi^2} q^2 P\left( q \right) \frac{j_0 \left( q s \right) -1}{\left( q s \right)^4}\, ,
\ee
which is equivalent to removing the variance of each of the divergent pieces
\be
\xi^{\rm div} \left( s, s_1 , \mu \right) = \sum_{\cal{O}\cal{O}'}\xi^{\rm div}_{\cal{O}\cal{O}'}(s,s_1,\mu)  
= \sum_{\cal{O}\cal{O}'}\tilde{\xi}^{\rm div}_{\cal{O}\cal{O}'} + \sum_{\cal{O}\cal{O}'} (\sigma^2)^{\rm div}_{\cal{O}\cal{O}'}\, ,
\ee
where the new functions $\tilde{\xi}^{\rm div}_{\cal{O}\cal{O}'}$ are defined by the replacement of $I^4_0$ with $\tilde I^4_0$.
We can now compute all the $\tilde{\xi}$'s, which are finite, but we are left with the problem of adding a number of infinite terms at the end. 
However, it turns out that the sum of the variance of the divergent pieces  is identically zero,
\begin{align}
\label{eq:k-4}
   \sum_{\cal{O}\cal{O}'} (\sigma^2)^{\rm div}_{\cal{O}\cal{O}'} =0\,,
\end{align}
once all the terms have been taken into account. Crucially the terms computed at the observer's position are necessary for the cancellation.

This result implies that the lower integration limit in $I_0^4(s)$ does not play any role once all possible correlations have been taken into account, or equivalently that the final result is not sensitive to the large scale gravitational potentials.
The cancellation of the IR-divergent terms in the variance of the galaxy density field however does not imply that the observed power spectrum does not contain terms scaling like $P(k)k^{-4}$.
Subtracting a constant with respect to $s$ as in \eq{eq:Itilde} indeed changes only the zero mode, $k=0$, of \avg{P_L(k)}. It is straightforward to see that for any finite mode $k>0$, the Hankel transform of $\tilde{I}_0^4$ and $I_0^4$ are the same\footnote{We used
\be \nonumber
\int \dd s s^2 j_0 \left( k s \right) \stackrel{(k>0)}{=}  - \left( \partial_k^2 + \frac{2}{k} \partial_k \right) \int \dd s j_0 \left( k s \right) = - \left( \partial_k^2 + \frac{2}{k} \partial_k \right) \frac{\pi}{2 k} = 0 \, .
\ee}

\bea
\label{eq:integral_I40}
4 \pi  \int \dd s s^2 s^4 \tilde I^4_0 \left( s \right) j_0 \left( k s \right) 
&=& 4 \pi  \int \dd s s^2 j_0 \left( k s \right)  \int \frac{\dd q}{2 \pi^2} q^2 P \left( q \right) \frac{j_0 \left( q s \right) -1}{q^4} \nonumber \\
&=&
k^{-2} \int \dd q q^{-2} P(q) \delta_D \left( q - k \right) - \frac{2}{\pi} \int \frac{\dd q}{q^2} P\left( q \right) \int \dd s  s^2 j_0 \left( k s \right) \nonumber \\
&\stackrel{(k>0)}{=}& P(k) k^{-4} \, .
\eea

We therefore conclude that the true observed power spectrum will exhibit $k^{-4}P(k)$-like behavior, which can however be exactly computed without any ad hoc procedure to tame the divergences.

The proof of \eq{eq:k-4} is long and tedious, and it is easily implemented with symbolic programming that can deal automatic with all the different pieces\footnote{A Mathematica notebook showing the cancellation is attached to the ArXiv submission.}. It is however instructive to look at a subset of terms, specifically the ones proportional to $f_{\rm evo}$, which cancels among themselves since $f_{\rm evo}$ is an arbitrary free parameter of the model. The derivation illustrates the physical ingredients required for the cancellations. The terms we are interested in are
\be 
\label{eq:evo_terms}
\Delta_{\rm evo} \equiv - f_{\rm evo } \left(
\HH_0 V_o + \Psi - \Psi_o + \ndv - \ndv_o -  \HH V + \int_\tau^{\tau_o} \left( \dot \Psi + \dot \Phi \right)\dd\tau' \right) \, .
\ee
By using \eq{eq:mapping_start} we can compute the correlation function
\be
\xi^{\Delta_{\rm evo}} \left( s_1 , s_2 , \cos\theta \right) = \int \frac{\dd q}{2 \pi^2} q^2 P \left( q \right)\int_0^{s_1}\dd \chi_1 \int_0^{s_2}\dd \chi_2  \mathcal{D}_{\rm evo} \left( q ,s_1, \chi_1\right) \mathcal{D}_{\rm evo} \left( q ,s_2, \chi_2\right) j_0 \left( q \chi \right)  
\ee
where $\mathcal{D}_{\rm evo}$ is the differential operator associated to \eq{eq:evo_terms}, through the mapping in \eq{eq:mapping_start}. 
In order to show that IR divergent contributions to the variance vanish we need to check that
\be \label{eq:limit_q4}
\lim_{q \rightarrow 0} q^4 \int_0^{s_1}\dd \chi_1 \int_0^{s_2}\dd \chi_2  \mathcal{D}_{\rm evo} \left( q ,s_1, \chi_1\right) \mathcal{D}_{\rm evo} \left( q ,s_2, \chi_2\right) j_0 \left( q \chi \right)   =0 \, .
\ee
A little algebra shows that this condition is satisfied only if 
\be
\int_0^{s_1}\dd\chi_1 \left(T_{\dot \Psi} (\chi_1) +T_{\dot \Phi} (\chi_1) \right)-\HH(s_1) T_V(s_1)+\HH_0 T_V(0)+T_\Psi(s_1)-T_\Psi(0) =0 \, ,
\ee
which by integrating over $\chi_1$ reduces to
\be
 T_\Phi \left( s_1 \right)  + \HH \left( s_1 \right) T_V \left( s_1 \right)  = T_\Phi \left( 0 \right) +\HH_0 T_V \left( 0 \right) = {\rm constant} \, .
\ee
In terms of the metric perturbation and the peculiar velocity this condition reads  
\be \label{eq_condition}
\Phi\left( k , \tau \right)  + \frac{\HH\left(  \tau \right)}{k}   v\left( k , \tau \right)  = {\rm constant} \qquad \text{for } \quad k \rightarrow 0 \, .
\ee
It is easy to show that this condition is satisfied by the Weinberg adiabatic mode~\cite{Weinberg:2003sw}, under the assumption that dark matter is the only clustering species. The existence of such adiabatic mode indeed guarantees that the long wavelength gravitational potential is unobservable and can always be reabsorbed with a change of coordinates.
In this case for the dark matter velocity we have 
\be
v_M = k \frac{\dot \Phi + \HH \Phi }{4 \pi G a^2  \bar \rho_M} =  k  \frac{ \dot \Phi + \HH \Phi}{\HH^2-3 \dot\HH}\, ,
\ee
where we have explicitly used $\Phi = \Psi$. Then by plugging this expression into \eq{eq_condition} and taking a time derivative we obtain a second order differential equation for $\Phi$
\be
\label{eq:diff_Weinberg}
\ddot \Phi  +3 \HH \dot \Phi+ \left( \HH^2+ 2 \dot \HH \right) \Phi = 0 \, ,
\ee
where we have used
\be
\dddot a= \frac{2}{a^2} \left( 2 a \dot a \ddot a -\dot a^3 \right) \, .
\ee
The differential equation in \eq{eq:diff_Weinberg} is precisely solved by the Weinberg adiabatic mode\footnote{In \eq{eq:Weinberg} we have neglected the decaying mode.}~\cite{Weinberg:2003sw,Mirbabayi:2014hda}
\be \label{eq:Weinberg}
\Psi=\Phi = C_1 \left( \frac{\HH}{a^2} \int_0^{\tau} a^2 d\tau' -1 \right) \, .
\ee
Therefore we conclude that cancellation of the IR divergences is a direct consequence of the existence of the Weinberg adiabatic mode as a solution of the Einstein field equations. 
We also recognize that \eq{eq_condition} agrees with the spatial curvature perturbation on co-moving spatial surface
\be
- \mathcal{R}_{co}\left( k , \tau \right)  = \Phi\left( k , \tau \right)  + \frac{\HH\left(  \tau \right)}{k}   v\left( k , \tau \right) 
\ee
which is conserved on super-Hubble scales for adiabatic perturbations in absence of anisotropic stresses.

In presence of local Primordial Non Gaussianities, the sensitivity of the two-point function to the long-wavelength gravitational potential is physical, and it cannot therefore be removed. In this case, the correlation between large and small scale modes is actually imprinted during inflation. 
On the other hand, it is possible to show that the IR divergences are solely due to the auto-correlation of the primordial gravitational potential, and are hence proportional to $f_{\rm NL}^2$. The terms involving one power of $f_{\rm NL}$ and a metric perturbation are all individually divergent in the IR, but once again their sum is finite. This result is a consequence of the fact that in the squeezed limit local PNG cannot be generated by gravitational evolution.

Finally we notice that from the expression derived in this section, \eq{eq:limit_q4}, directly implies
\be
\lim_{q \rightarrow 0} q^3 \int_0^{s_1}\dd \chi_1 \int_0^{s_2}\dd \chi_2  \mathcal{D}_{\rm evo} \left( q ,s_1, \chi_1\right) \mathcal{D}_{\rm evo} \left( q ,s_2, \chi_2\right) j_0 \left( q \chi \right)   =0   \, .
\ee

\subsection{The $1/q^2$ IR sensitivity}
The terms involving one power of the metric potentials are not IR sensitive in a $\Lambda$CDM Universe with parameters as determined by the Planck satellite~\cite{Aghanim:2018eyx}.
Nevertheless, in the limit $q\rightarrow0$ we expect the gravitational potentials to disappear from the correlation function as shown in the previous section. This is an important consistency check of the theory, because, if the cancellation is a property of General Relativity and of Gaussian and adiabatic initial conditions, then long-wavelength modes are unobservable regardless of the actual value of cosmological parameters.
Following the example in the previous section we show this is the case for the contribution in \eq{eq:evo_terms}, while a complete treatment can be easily obtained through symbolic programming.

Inspection of \eq{eq:evo_terms}, reveals that we need to consider two set of terms to show the cancellation of a single metric potential in the large scale limit. Indeed we do not only have the contribution induced by the peculiar velocities, \ie proportional to the gradient of the gravitational potential, but also the terms proportional to two powers of the gravitational potential combined with the expansion of the spherical Bessel at the order $q^2$.  In terms of the transfer function $T_\Psi$, $T_{\dot \Psi}$ and $T_V$ we arrive to
\bea
\label{eq:q2}
&&  \hspace{-3cm} \lim_{q \rightarrow 0} q^2 \int_0^{s_1}\dd \chi_1 \int_0^{s_2}\dd \chi_2  \mathcal{D}_{\rm evo} \left( q ,s_1, \chi_1\right) \mathcal{D}_{\rm evo} \left( q ,s_2, \chi_2\right) j_0 \left( q \chi \right)   
\nonumber \\
&=& \left( s_1 s_2 \right)^{-1} \frac{1}{6} \int_0^{s_1}\dd \chi_1 \int_0^{s_2}\dd \chi_2 
\nonumber \\
&&
\left\{ 2 \chi_1^2 s_1 T_{\dot \Psi}(\chi_1) (T_{\Psi}(0)-\HH_0 T_V(0))-4 \Delta\chi^2 s_1 s_2 T_{\dot \Psi}(\chi_1) T_{\dot \Psi}(\chi_2)
\right. 
\nonumber \\
&&
+4 s_1 T_{\dot \Psi}(\chi_1) T_V(s_2) (\chi_1 y-s_2)-4 \chi_1 y s_1 T_V(0) T_{\dot \Psi}(\chi_1)
\nonumber \\
&&
-2 \Delta\chi_1^2 s_1 T_{\dot \Psi}(\chi_1) (T_{\Psi}(s_2)-\HH(s_2) T_V(s_2))
\nonumber \\
&&
+2 \chi_2^2 s_2 T_{\dot \Psi}(\chi_2) (T_{\Psi}(0)-\HH_0 T_V(0))
\nonumber \\
&&
+4 s_2 T_{\dot \Psi}(\chi_2) T_V(s_1) (\chi_2 y-s_1)-4 \chi_2 y s_2 T_V(0) T_{\dot \Psi}(\chi_2)
\nonumber \\
&&
-2 \Delta\chi_2^2 s_2 T_{\dot \Psi}(\chi_2) (T_{\Psi}(s_1)-\HH(s_1) T_V(s_1))
\nonumber \\
&&
+2 T_V(s_1) (y s_1 T_V(0) \HH(s_1)-T_V(0) (y+\HH_0 s_1)+s_1 T_{\Psi}(0))
\nonumber \\
&&
+2 T_V(s_2) (y s_2 T_V(0) \HH(s_2)-T_V(0) (y+\HH_0 s_2)+s_2 T_{\Psi}(0))
\nonumber \\
&&
+2 T_V(s_2) (T_V(s_1) (\HH(s_1) (s_2-y s_1)+\HH(s_2) (s_1-y s_2)+y)
\nonumber \\
&&
+T_{\Psi}(s_1) (y s_1-s_2))-2 T_V(s_1) T_{\Psi}(s_2) (s_1-y s_2)
\nonumber \\
&&
-2 y s_1 T_V(0) T_{\Psi}(s_1)-2 y s_2 T_V(0) T_{\Psi}(s_2)+2 y T_V(0)^2
\nonumber \\
&&
+s_1^2 (T_{\Psi}(0)-\HH_0 T_V(0)) (T_{\Psi}(s_1)-\HH(s_1) T_V(s_1))
\nonumber \\
&&
+s_2^2 (T_{\Psi}(0)-\HH_0 T_V(0)) (T_{\Psi}(s_2)-\HH(s_2) T_V(s_2))
\nonumber \\
 && \left.
-s^2 (T_{\Psi}(s_1)-\HH(s_1) T_V(s_1)) (T_{\Psi}(s_2)-\HH(s_2) T_V(s_2))
\right\} \, ,
\eea
where we have introduced $y=\cos\theta$.
Now combining the Euler equation
\be
\dot T_V + \HH T_V= T_\Psi
\ee
and the condition in \eq{eq_condition}, which we remind is satisfied Weinberg the adiabatic mode, we obtain
\be
T_{\Psi} = \frac{1}{2} \left( T_\Psi \left( 0 \right) + \HH_0 T_V \left( 0 \right) +\dot { T}_V
\right) 
\ee
or equivalently
\be
\int_0^{s_1}\dd \chi_1 T_\Psi \left( \chi_1 \right) = \frac{s_1}{2} \left[ T_\Psi \left( 0 \right) + \HH_0 T_V \left( 0 \right) + \frac{1}{2} \left( T_V \left( 0\right) - T_V \left( s_1 \right) \right) \right] \, . 
\ee
After some integration by parts and using the Friedman Equation $
\dot \HH = \HH^2 - \frac{3 \HH_0^2 \Omega_{M 0}}{2 a}
$,
we obtain the desired cancellation
\be
\lim_{q \rightarrow 0} q^2 \int_0^{s_1}\dd \chi_1 \int_0^{s_2}\dd \chi_2  \mathcal{D}_{\rm evo} \left( q ,s_1, \chi_1\right) \mathcal{D}_{\rm evo} \left( q ,s_2, \chi_2\right) j_0 \left( q \chi \right)   =0   
\, ,
 \ee
 and as a consequence also that
 \be
\lim_{q \rightarrow 0} q \int_0^{s_1}\dd \chi_1 \int_0^{s_2}\dd \chi_2  \mathcal{D}_{\rm evo} \left( q ,s_1, \chi_1\right) \mathcal{D}_{\rm evo} \left( q ,s_2, \chi_2\right) j_0 \left( q \chi \right)   =0    \, .
 \ee

Similarly to the discussion of the divergent terms in the previous section, the cancellation only affects the zero mode of the power spectrum, hence GR contributions proportional to $P(k)k^{-2}$ are present at any finite scale.

To summarise this Section, we found that the existence of the Weinberg adiabatic mode removes the IR sensitivity of the correlation function. The presence of the adiabatic mode is guaranteed for Gaussian and adiabatic initial conditions and by the diffeomorphism invariance of General Relativity. Notice all three conditions are required. 
We already discussed how local PNG induce correlation between long and short wavelength modes. If the initial conditions contained some amount of isocuvature perturbations or new long range forces are present in the dark sector, then the comoving curvature perturbation is not constant on super-horizon scales and the correlation function could exhibit a weak IR dependence. 

While our calculation addresses the specific problem of the computation of the galaxy correlation function and power spectrum, similar arguments about the effects of long wavelength gravitational potentials, or better said lack thereof, on cosmological observables have long been known. For the late Universe, they usually exploit the Consistency Relations for LSS~\cite{Creminelli2013,Peloso2013,Kehagias2013}, which are valid under the same conditions required by the cancellation we showed above. For an application of the Consistency Relations to the relativistic bispectrum see~\cite{Kehagias:2015tda}.
Our result also implies that GR effects are not coupled to any super sample value of the gravitational potential and its spatial derivative, as it would be the case in presence of local PNG \cite{Castorina2020}.

Our findings are in disagreement with the recent work in \cite{Grimm:2020ays}. They assume that the observed power spectrum constructed from squaring the Fourier transform of the galaxy overdensity field is diagonal, \ie the field is homogeneous and isotropic. But as we discussed in the introduction, RSD and GR effects break translational and rotational invariance: the Yamamoto estimator is a function of a single $k$ mode because the other one has been integrated out. The power spectrum discussed in \cite{Grimm:2020ays} is therefore not directly related to any observable.
Where we agree with \cite{Grimm:2020ays} is in the calculation of the variance of the galaxy overdensity field, which does not receive contribution from the gravitational potential and its first derivatives. As discussed above we clarified the necessary physical conditions for this cancellation, which requires more than just the equivalence principle to hold as suggested by  \cite{Grimm:2020ays}.
The full two-point correlation function of the galaxy density field and any general estimator for the power spectrum like the one~\eq{eq:def_estimator}, are complicated average of all possible lines of sight for each pair of galaxies and should not be compared with the variance of the field itself, which is a function of a single line of sight.

Finally, one could argue that the any IR sensitivity of the model adopted here would be anyway removed by the presence of an Integral Constraint (IC). This could be the case, however it is important to check that the relativistic expression for the galaxy number counts provide a finite answer regardless of the way the mean number density of galaxy is measured. If one wants to resort to the IC, the latter should be included at the level of the integrand in Eq.~(\ref{eq:estimator_standard}), carefully including the effect of the window function. In our approach we removed the IR sensitivity in both Eq.~(\ref{eq:estimator_standard}) and in the expression of the IC, which can be therefore discarded for most practical applications. 

The method described in this work to remove the IR divergences is also the only one that can deal with observations in which the mean is not known or cannot be measured. This is for instance the case of the excess brightness temperature of the 21 cm line, or any intensity line, measured with radio arrays in interefometric mode. With interferometers we directly have access to the fluctuations in Fourier Space, and there is no IC to subtract to the measurements of the power spectrum.
%%%%%%%%%%%%%%%%%
\section{Relativistic effects in the galaxy power spectrum}
\label{sec:GR_effects_Pk}
Having resolved the issue of IR divergences in the correlation function we can now safely proceed to compute the prediction for the observed galaxy power spectrum multipoles. Our calculation of the correlation function includes the observer's terms, in auto-correlation with themselves and in cross-correlations with the source terms. 
Given the gauge invariant galaxy number counts, \eq{eq:Delta_rel}, we still have the freedom to perform a Lorentz boost to a preferred reference frame. In particular we are allowed to move to the CMB rest frame, and remove the terms proportional to the observer's velocity. While this operation is harmless in the theoretical prediction, some care should be used when expressing the measurement in the CMB or any other frame~\cite{Kaiser:1987qv,Hamilton:1997zq}. Alternatively one can keep these terms and try to measure the induced dipole to test the Copernican Principle. We provide expressions for this case in Appendix~\ref{sec:observer_vel_appendix}.

\begin{figure}[t!]
\begin{center}
\includegraphics[width=1.09\textwidth]{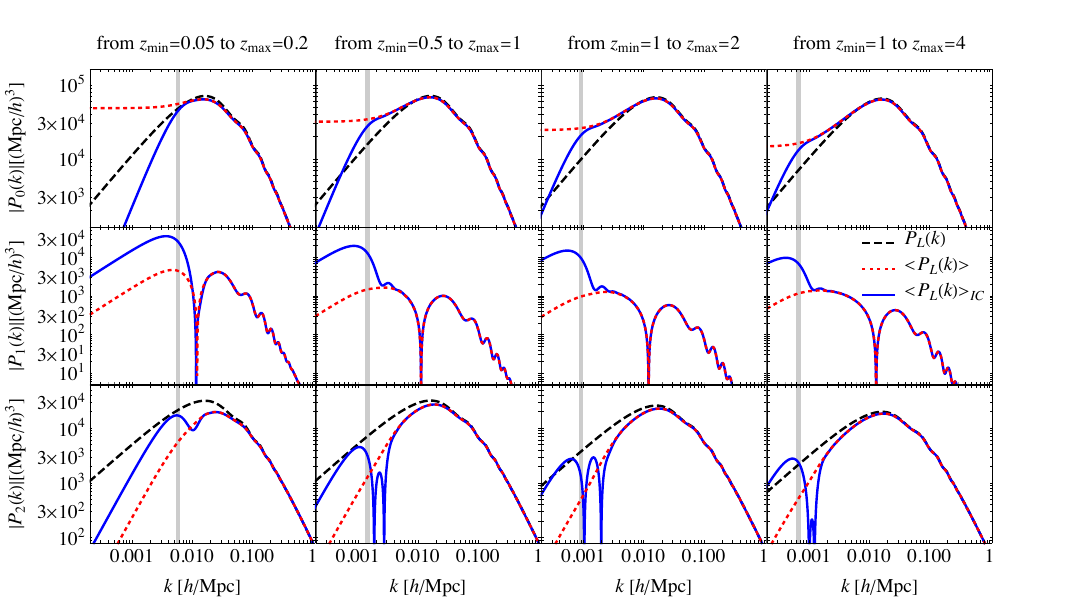}
\vspace{-0.75cm}
\caption{We plot the monopole (top panels), the dipole (middle panels) and the quadrupole (bottom panels) of the power spectrum for four different survey volumes. Different colors denote: the flat-sky theoretical power spectrum without any window function (dashed black), the full-sky power spectrum convoluted with the survey geometry (dotted red), including also the integral constraint (solid blue). The vertical lines correspond to the largest scales probed by the survey, \ie $2 \pi /\left( 2 r (z_{\rm max}) \sin\theta_{\rm max} \right)$.}
\label{fig:RSD_pk}
\end{center}
\end{figure}

The other required model ingredients are the multipoles of the window functions $F_\ell(s,s_1)$, and the Fourier space windows functions $Q_L(k)$. While our expressions are fully general, we choose a rather simple survey geometry, whose properties can be analytically calculated. We thus assume that the window function has an azimuthal symmetry around an axes, for instance a spherical cap with a maximum opening angle $\theta_{
\rm max}$. We fix this angle by imposing the hypothetical survey covers one third of the sky, $f_{\rm sky} = 1/3$. This assumption allows us to sum over all values of $\ell$ and $\ell_1$ in \eq{eq:estimator_standard}. We also assume that the radial selection function and angular footprint can be separated, which might not always be the case if different parts of the sky have large depth variations. Finally we assume a constant selection functions within a certain redshift range. We consider four redshift bins, $[z_{\rm min} = 0.05, z_{\rm max} =0.2]$, [0.5,1], [1,2] and [1,4]. For these choices of window function, the convolution with the correlation function further simplifies, as discussed in Appendix~\ref{sec:Azymuthal_window}.
Figure \ref{fig:window} shows the multipoles $Q_L(s)$ and $Q_L(k)$ in our four reference surveys, for $L=0,1,2$.
The largest pair separation for an
azimuthally symmetric survey geometry is given by $2 r (z_{\rm max}) \sin\theta_{\rm max}$, where $\theta_{\rm max}= \cos^{-1}\left( 1 - 2 f_{\rm sky}\right)$. We see that the scale $k \sim 2 \pi /\left( 2 r (z_{\rm max}) \sin\theta_{\rm max} \right)  $ indicates the boundary of the survey volume in Fourier space and we will display it in the other figures as well.
The effect of the window function is quite dramatic at low $k$, as one can see in Fig.~\ref{fig:RSD_pk}. The black dashed lines show the flat-sky prediction for the multipoles of the power spectrum without the window function, while the blue and red ones display the convolution with the window function, with and without the Integral Constraint respectively. 
Notice that our asymmetric choice of LOS generates a power spectrum dipole, as shown in the second row of Fig.~\ref{fig:RSD_pk}.

\begin{figure}[t]
\begin{center}
\includegraphics[width=1.09\textwidth]{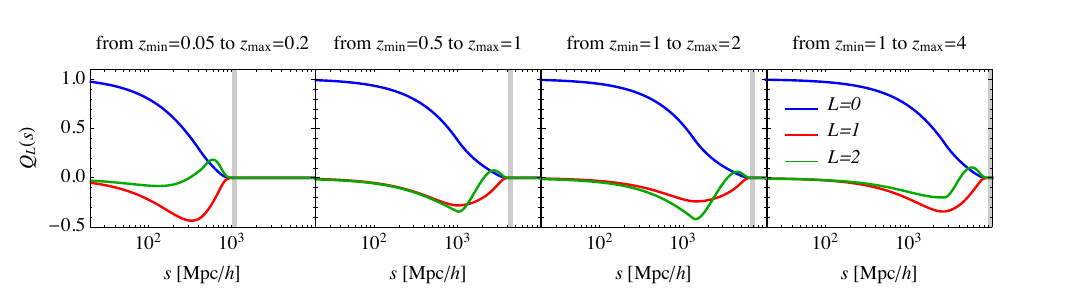} \\
\vspace{-0.5cm}
\includegraphics[width=1.09\textwidth]{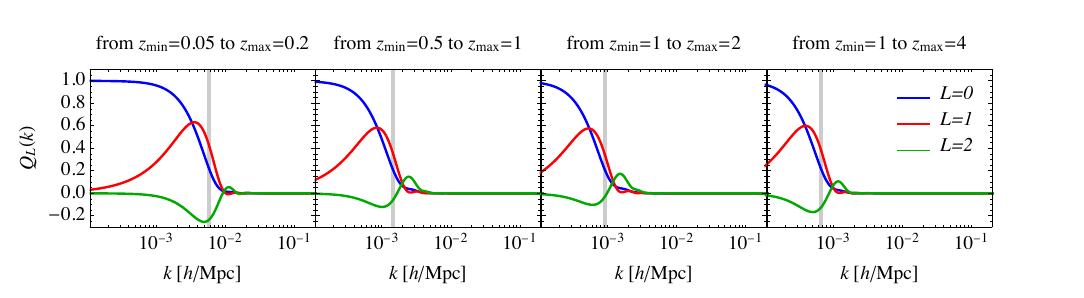}
\vspace{-0.75cm}
\caption{
We show the multipoles of the window functions considered in the manuscript $Q_L\left( s \right)$ and its Hankel transforms $Q_L\left( k \right)$ for the monopole ($L=0$), the dipole ($L=1$) and the quadrupole ($L=2$), normalized to $Q_0 \left( k \rightarrow 0 \right) =1$. The window function is defined as $\phi \left( \vs_1 \right) = \phi \left(s_1 \right) W \left( \hat \vs_1 \right)= \Theta\left( s_1 -s_{\rm min} \right) \Theta\left( s_{\rm max} -s_1 \right) \Theta\left( \theta_{\rm max}- \theta_1 \right)$, where $\theta_{\rm max}= \cos^{-1}\left( 1 - 2 f_{\rm sky}\right)$ and we set $f_{\rm sky}=1/3$. The vertical lines correspond to the largest scales probed by the survey, \ie $ 2 r (z_{\rm max}) \sin\theta_{\rm max} $ in real space and $2 \pi /\left( 2 r (z_{\rm max}) \sin\theta_{\rm max} \right)$ in Fourier space.}
\label{fig:window}
\end{center}
\end{figure}

%%%%%%%%%%%%%%%%%%%%
%%%%%%%%%%%%%%%%%%%%

\subsection{Density and RSD}
\label{subsec:densityRSD}

In the small angle limit the standard density and RSD correlation function and power spectrum correspond to the plane parallel limit of Kaiser~\cite{Kaiser:1987qv}. For wide area surveys, like DESI or Euclid, deviations from the plane parallel regime take the name of wide-angle effects.  The amplitude of these terms is controlled by $k s_{\rm eff}$, where $s_{\rm eff}$ is the comoving distance to the effective redshift, as opposed to $k/\cal{H}$ for the other relativistic effects. Therefore for surveys of the size of the Hubble horizon, $s_{\rm eff} \sim \HH^{-1}$, these effects are comparable, while for smaller surveys wide-angle effects are more important.

To derive the correlation function induced by density and redshift perturbations, \ie what we denote by the standard Newtonian contribution to galaxy clustering, we follow \eq{eq:correlation_differential}, 
\bea
&& \hspace{-1cm}\xi^{\rm newt} \left( s_1 , s_2 , \hat \vs_1 \cdot \hat \vs_2 \right)  = \langle \left( b_1 D_m+ \HH^{-1} \partial_r  \ndv  \right) \left(  \vs_1 \right) \left(b_1 D_m+ \HH^{-1} \partial_r  \ndv \right) \left( \vs_2 \right)  \rangle =
\nonumber \\
&=& \int \frac{\d q}{2 \pi^2} q^2  P\left( q \right) D_1 \left( s_1 \right) D_1 \left( s_2 \right) \left( b_1 \left( s_1 \right)  -f \left( s_1 \right) \partial^2_{q s_1} \right)  \left( b_1\left( s_2 \right) - f\left( s_2 \right)  \partial^2_{q s_2} \right) j_0 \left( q s \right) 
\nonumber \\
&=&
\sum_{i=0,2,4} J_i \left( s_1, s_2 ,\cos\theta \right) I^0_i \left( s \right) D_1\left( s_1 \right)  D_1\left( s_2 \right) \, ,
\eea
where
\bea
J_0 \left( s_1, s_2 , \cos\theta \right) &=& \frac{1}{15}  \left(f(s_1) \left(5 b_1(s_2)+2 {\cos^2\theta} f(s_2)+f(s_2)\right)+5 b_1(s_1) (3 b_1(s_2)+f(s_2))\right)\, , \nonumber
 \\
 && \\
 %%%%%%%%%%%%%%%%%%
J_2 \left( s_1, s_2 , \cos\theta \right) &=& -\frac{1}{21 s^2} 
\left\{
\left(7 b_1(s_2) f(s_1) \left(\left(3 {\cos^2\theta}-1\right) s_2^2-4 {\cos\theta} s_1 s_2+2 s_1^2\right)
\right. \right. \nonumber \\
 &&
+f(s_2) \left(7 b_1(s_1) \left(\left(3 {\cos^2\theta}-1\right) s_1^2-4 {\cos\theta} s_1 s_2+2 s_2^2\right)
\right. \nonumber \\
 &&\left. \left. \left.
+f(s_1) \left(\left(11 {\cos^2\theta}+1\right) s_1^2-4 {\cos\theta} \left({\cos^2\theta}+5\right) s_1 s_2
\right. \right. \right. \right.
\nonumber \\
 &&\left. \left. \left. \left.
 \qquad \qquad 
+\left(11 {\cos^2\theta}+1\right) s_2^2\right)\right)\right) \right\} \, ,
 \\
  %%%%%%%%%%%%%%%%%%
J_4 \left( s_1, s_2 , \cos\theta \right) &=& \frac{f(s_1) f(s_2)}{560 s^4 s_1^2 s_2^2} \left\{4 \left(30 {\cos^2\theta}-19\right) s_1^6 s_2^2-8 {\cos\theta} \left(4 {\cos^2\theta}+39\right) s_1^5 s_2^3
\right. \nonumber \\
 &&\left.
+18 \left(8 {\cos^2\theta}+23\right) s_1^4 s_2^4-8 {\cos\theta} \left(4 {\cos^2\theta}+39\right) s_1^3 s_2^5
\right. \nonumber \\
 &&\left.
+4 \left(30 {\cos^2\theta}-19\right) s_1^2 s_2^6+24 {\cos\theta} s_1^7 s_2+24 {\cos\theta} s_1 s_2^7+3 s^8-3 s_1^8-3 s_2^8\right\}\, .
\nonumber \\
&& 
\eea
To use \eq{eq:estimator_standard} we need to adopt the end-point line-of-sight convention 
\be \label{eq:end-point}
s_2 = \sqrt{s_1^2 + s^2 + 2 s_1 s \mu } \qquad \text{and} \qquad \cos \theta = \mu \frac{s}{s_2} + \frac{s_1}{s_2} \, .
\ee

Under the effective redshift approximation, the expression for the correlation function of density + RSD takes a simple, but long, analytical form. We have
\begin{align}
  \xi^{\rm newt}(s,z_{\rm eff}, \mu) &= \frac{D_1(z_{\rm eff})^2}{105 \left(1+x^2+2 x \mu\right)} \notag  \\
  \times&  \Bigg\{ 7 I_0^0(s) \Big[5 b_1 (3 b_1+f) \left(2 \mu  x+x^2+1\right)+5 b_1 f\left(2 \mu  x+x^2+1\right)  \notag\\
   & + f^2 (x (2 \mu  (\mu  x+3)+x)+3)\Big]  \notag\\
   &  -5  I_2^0(s) \Big[7 b_1 f \left(-1+3 \mu ^2\right) \left(1+2 \mu  x+x^2\right) +f [(7 b_1+6 f) \notag\\
   &  \left(-1+3 \mu ^2\right)+2 \mu  \left(14 b_1 +3 f+9 f \mu ^2\right) x+\left(14b_1+f+11 f \mu ^2\right) x^2]\Big] \notag\\
   & + 3 I_4^0(s) f^2 \left[3+35 \mu ^4-24 \mu x+40 \mu ^3 x-4 x^2+6 \mu ^2 \left(-5+2
   x^2\right)\right]\Bigg\}\,  ,
\end{align}
where $x = s/r(z_{\rm eff})$  and $b_1$ and $f$ are evaluated at the effective redshift $z_{\rm eff}$. The above expression can then be projected into multipoles.
Notice that in linear theory all multipoles, even and odd ones, are non zero, and only in the plane-parallel limit when $x\rightarrow0$ we recover the standard expressions in~\cite{Kaiser:1987qv}.
\begin{figure}[t]
\begin{center}
\includegraphics[width=0.49\textwidth]{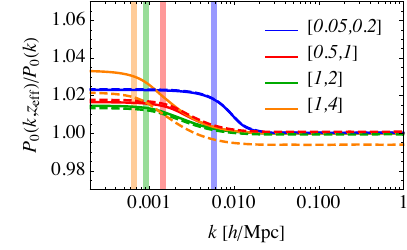}
\includegraphics[width=0.49\textwidth]{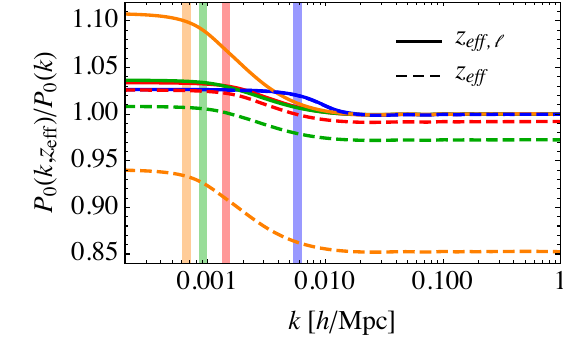}
\\
\includegraphics[width=0.49\textwidth]{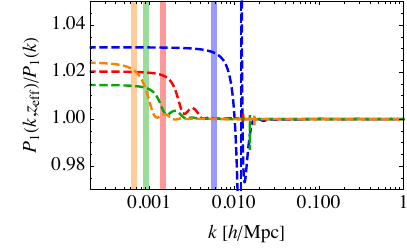}
\includegraphics[width=0.49\textwidth]{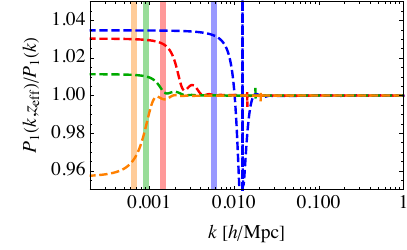}
\\
\includegraphics[width=0.49\textwidth]{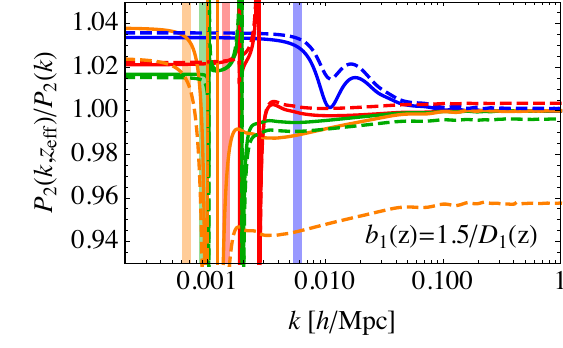}
\includegraphics[width=0.49\textwidth]{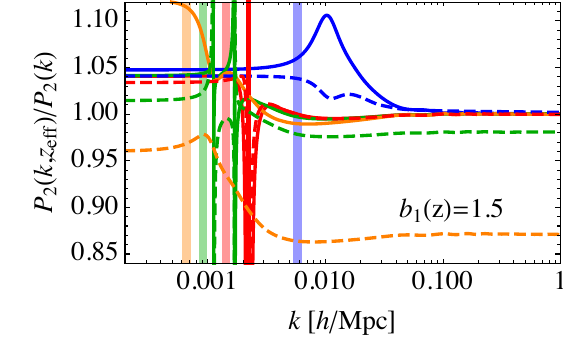}
\vspace{-0.25cm}
\caption{
We plot the ratio between the power spectrum within the effective redshift approximation and the full redshift dependence for the monopole (top panels),  the dipole (middle panels) and the quadrupole (bottom panels). The different colors are associated to the following constant redshift selection functions: $[0.05,0.2]$ blue, $[0.5,1]$ red, $[1,2]$ green and $[1,4]$ orange.
Solid lines refer to the theoretical effective redshift defined in \eq{eq:def_zeff_th}, while dashed lines to the volume weighted mean redshift, see \eq{eq:def_zeff}.
The vertical lines correspond to the largest scales probed by the survey, \ie $2 \pi /\left( 2 r (z_{\rm max}) \sin\theta_{\rm max} \right)$.
In the left panels we consider $b_1= 1.5/D_1 \left( z \right) $, while on the right panels $b_1 =1.5$.}
\label{fig:eff_redshift}
\end{center}
\end{figure}
The accuracy of the effective approximation is shown in \fig{fig:eff_redshift}, where we plot the ratio between the full prediction, \ie integrated over $s_1$, and the power spectrum evaluated at the effective redshift, for the monopole, dipole and quadrupole. The prediction includes only the density plus RSD piece. 
For the left panels we assumed the linear bias scales with the growth factor, $D(z)b_1(z)=1.5$, while on the right ones we fixed $b_1(z)=1.5$. Different colors show different surveys, and the vertical bars display the largest scales measured in each of them. Continuous lines indicate the prediction evaluated at the $z_{\rm eff,\,\ell}$ (see \eq{eq:def_zeff_th}), which by definition are unbiased at sufficiently high $k$, while dashed ones use the global $z_{\rm eff}$ (see \eq{eq:def_zeff}). Notice that for the dipole we only show the global $z_{\rm eff}$ since in the plane parallel limit there is no intrinsic dipole.
This figure shows that the effective redshift approximation with the $z_{\rm eff,\,\ell}$'s is in general a very good approximation, better than a \% in all cases except the more unrealistic scenario of a single redshift bin galaxy survey between $z_{\rm min}=1$ and $z_{\rm max}=4$. The global $z_{\rm eff}$ performs slightly worse, and it is less accurate for constant galaxy bias. 
We also notice that the error introduced by the effective redshift approximation is usually multiplicative, and it can therefore be absorbed in a redefinition of the in principle unknown galaxy bias.

In real data the accuracy of the effective redshift prediction will depend on other observational effects, like completeness, that can be trivially included as additional redshift weights in our theoretical predictions.

\subsubsection{Full sky and $s/s_1$ expansion}
To obtain the estimated power spectrum we need to convolve the expression of the correlation function with the multipoles of the window function. 
Because of the finite survey volume, rather than computing the full $\xi_\ell(s,s_1)$, a more practical approach for the convolution with the window function is to expand the multipoles in powers of $s/s_1$ 
\be
\xi_\ell \left( s_1 , s  \right) = \sum_j \left(  \frac{s}{s_1}\right)^j \xi_\ell^{(j)} \left(  s \right)  \, .
\ee
Notice that in the above expression the $\xi_{\ell}^{(j)}$ might still explicit depend on $s_1$ to account for the redshift dependence of the growth factors.

\begin{figure}[t]
\begin{center}
\includegraphics[width=1.09\textwidth]{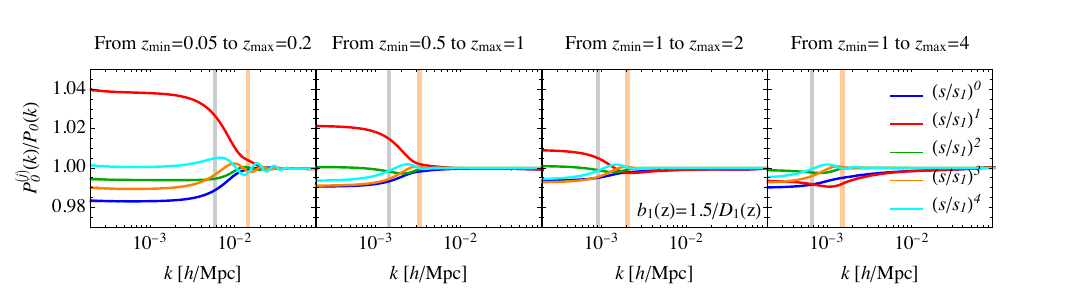}\\
\vspace{-0.5cm}
\includegraphics[width=1.09\textwidth]{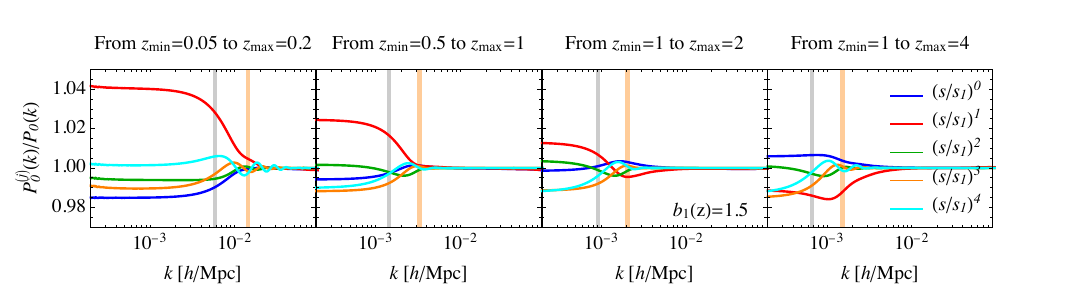}
\vspace{-0.75cm}
\caption{
We plot the ratio between the monopole of power spectrum expanded up to different order in the wide-angle parameter $s/s_1$ and the full-sky power spectrum.
The panels show the four different surveys with increasing redshift coverage from left to right. The vertical orange line indicates $k=2 \pi/r(z_{\rm eff})$, while the gray line refers to the largest scales within the survey geometry. In the top panels we consider $b_1= 1.5/D_1 \left( z \right) $, while on the bottom panels $b_1 =1.5$.
}
\label{fig:ful_vs_WA_monopole}
\end{center}
\end{figure}
The series expansion is certainly a bad approximation close to the survey boundary where the separation $s$ could approach the distance to the pair, and it eventually diverges, but the finite size of the window function addresses many of these shortcomings. In Fig.~\ref{fig:ful_vs_WA_monopole} we show the ratio between the monopole of the predicted power spectrum up to a given order in the series expansion, $\hat{P}^{(j)}_L(k)$, and the full calculation. In all four cases considered we assume a sky fraction $f_{\rm sky} = 1/3$ and vary the redshift range of the measurements. We find that the second order expansion in $s/s_1$, usually employed in the literature~\cite{Castorina:2017inr,Beutler:2018vpe}, is always a very good approximation up to the survey boundary, indicated by the thick vertical gray line in the panels. The difference of a few \% is always well below the expected cosmic variance for these hypothetical surveys. Notice how the presence of the window function makes the flat-sky prediction, the blue lines at $\mathcal{O}(s/s_1)^0$, very accurate.

\begin{figure}[!ht]
\begin{center}
\includegraphics[width=1.09\textwidth]{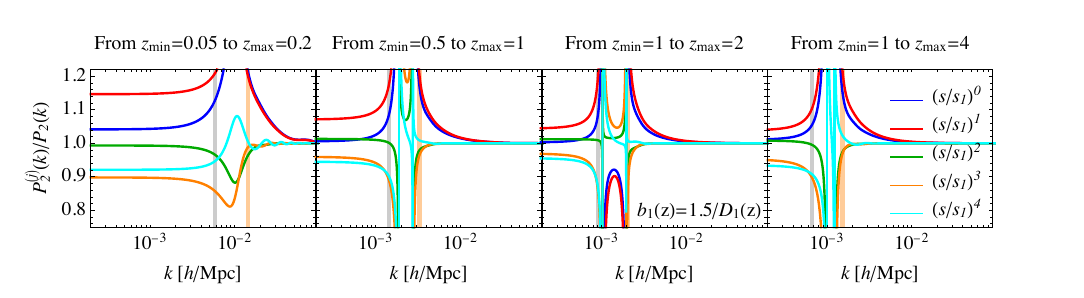}\\
  \vspace{-0.5cm}
\includegraphics[width=1.09\textwidth]{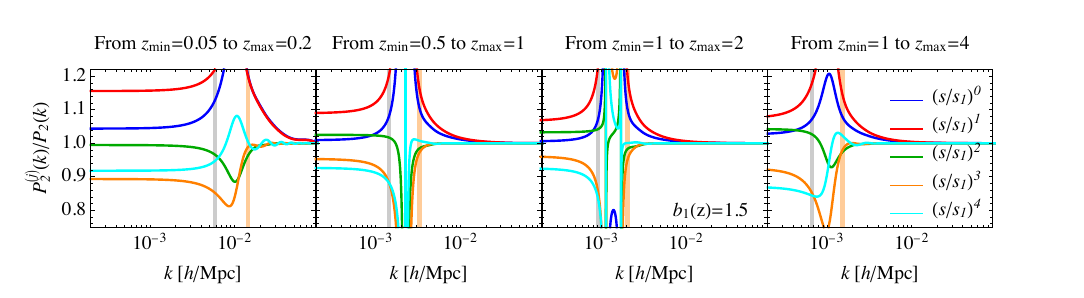}
  \vspace{-0.75cm}
\caption{
We plot the ratio between the quadrupole of the power spectrum expanded up to different orders in the wide-angle parameter $s/s_1$ and the full-sky power spectrum.
The four panels show survey geometry with different deepness in redshift. The vertical orange line indicates $k=2 \pi/r(z_{\rm eff})$, while the gray line refers to the largest scales within the survey geometry. In the top panels we consider $b_1= 1.5/D_1 \left( z \right) $, while on the bottom panels $b_1 =1.5$.
}
\label{fig:ful_vs_WA_quadrupole}
\end{center}
\end{figure}
The same set of plots is shown for the quadrupole in Fig.~\ref{fig:ful_vs_WA_quadrupole}.
As expected the series expansion performs slightly worse than for the monopole, but it should be kept in mind that the cosmic variance increases accordingly. 
The perturbative expansion in $s/s_1$ breaks down at scale larger than $k \sim 2 \pi/r_{\rm eff}$, shown as the orange bar in the Figure. Indeed this scale corresponds to $s/s_1 \sim 1$ in configuration space.

 \begin{figure}[!ht]
 \begin{center}
\includegraphics[width=1.09\textwidth]{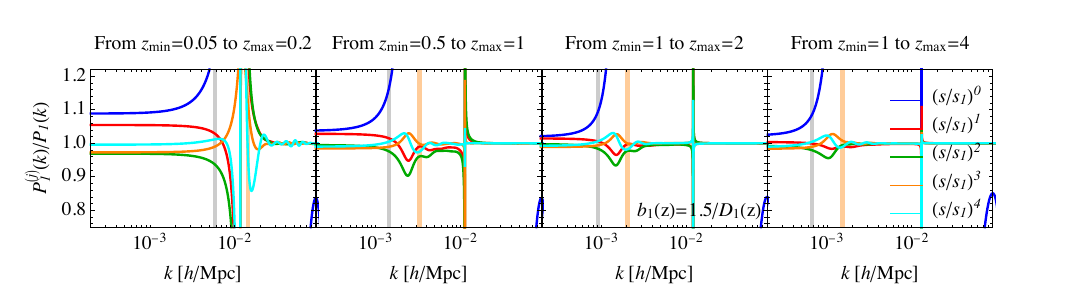} \\
  \vspace{-0.5cm}
\includegraphics[width=1.09\textwidth]{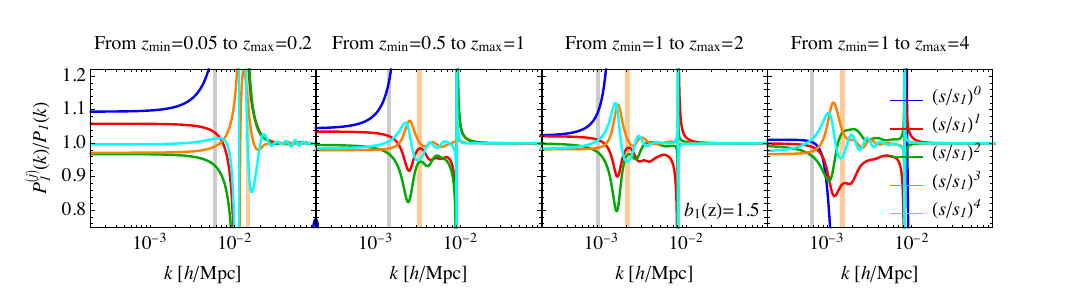}
  \vspace{-0.75cm}
\caption{
We plot the ratio between the dipole of power spectrum expanded up to different orders in the wide-angle parameter $s/s_1$ and the full-sky power spectrum.
The four panels show survey geometry with different deepness in redshift. The vertical orange line indicates $k=2 \pi/r(z_{\rm eff})$, while the gray line refers to the largest scales within the survey geometry. In the top panels we consider $b_1= 1.5/D_1 \left( z \right) $, while on the bottom panels $b_1 =1.5$.}
\label{fig:ful_vs_WA_dipole}
\end{center}
\end{figure}
For the dipole (Fig.~\ref{fig:ful_vs_WA_dipole}) the perturbative expansion starts to converge at $\mathcal{O}\left( s/s_1\right)$. Because the dipole of the correlation function vanishes for a single tracer, $P_1\left( k \right) $ in the plane-parallel approximation receives contributions only from the coupling between the window function and other multipoles.

\subsection{Lensing Magnification}

At fixed pair separation $s$ between two observed galaxies, terms like lensing magnification or ISW  measure the correlation between points along the lines of sight of the two galaxies. The "non-locality" of such contributions makes harder to  understand to which value of $k$ in the power spectrum they mostly contribute. In our formalism, however, computing the average of the observed power spectrum requires only the evaluation of the three-dimensional correlation function multipoles. We compute $\xi(s_1,s_2,\cos \theta)$ or $\xi(s,s_1,\mu)$ using FFTs, and  then Monte-Carlo or quadrature methods for the integrals along the LOS and the projection into multipoles. 

In this section we focus on lensing magnification, which enters the power spectrum at the same order in derivatives as the standard density and RSD terms. As mentioned before, we set $s_b=0$ as its value changes only the normalization of the lensing contribution to the power spectrum.

\begin{figure}[t]
 \begin{center}
  \includegraphics[width=1.09\textwidth]{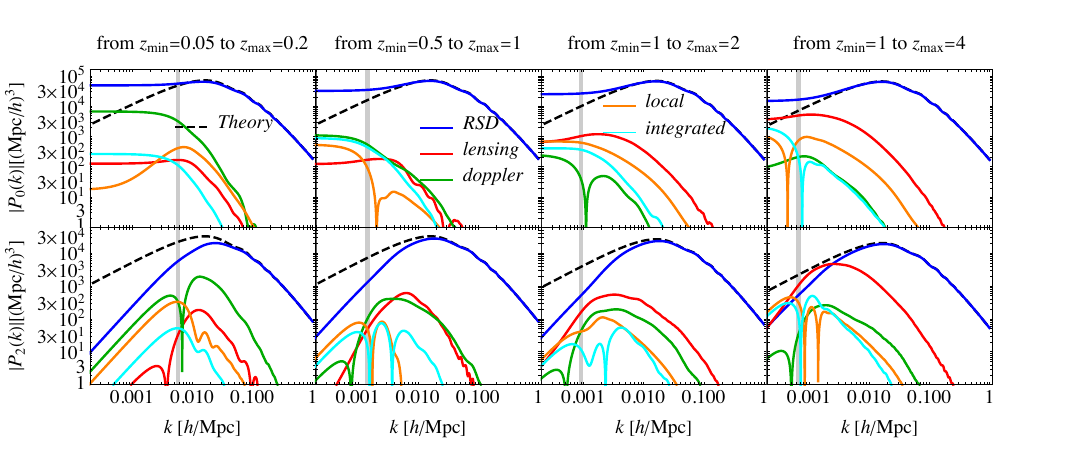}
  \vspace{-0.75cm}
\caption{
We plot the contributions of all the relativistic effects to the monopole (top panels) and quadrupole (bottom panels) of the power spectrum. Each relativistic effect includes also the correlation with the other effects listed above it in the legend. We remark that we are not considering the integral constraints in this figure, since we want to highlight the contribution of the single relativistic effects.
The vertical lines correspond to the largest scales probed by the survey, \ie $2 \pi /\left( 2 r (z_{\rm max}) \sin\theta_{\rm max} \right)$.
}
\label{fig:GR_effects_NOIC}
 \end{center}
\end{figure}

Fig.~\ref{fig:GR_effects_NOIC} shows in red the contribution of lensing magnification to the monopole and quadrupole of the power spectrum, which can be compared to the RSD term (including wide angle effects) in blue. Both curves include the effect of the window function, but not the integral constraint. For reference the flat-sky linear theory prediction in absence of a window function is shown as the dashed black line.
Fig.~\ref{fig:GR_effects_ratio} instead shows, again in red, the relative difference between the RSD power spectrum and the RSD plus lensing power spectrum. Because we are adding the different contributions we can now consistently subtract the integral constraint. 
As expected lensing magnification is a tiny effect for low redshift surveys with a small extension in redshift, as displayed in the leftmost panels. In this case, for both the monopole and the quadrupole  we find lensing magnification leads to sub-\% changes on the total power spectrum. 
Moving to higher redshift and wider redshift ranges the effect of lensing increases, reaching tens of percent for the monopole and $\mathcal{O}(1)$ for the quadrupole in our most unrealistic setup of a single redshift bin between $z=1$ and $z=4$ over one third of the sky.  We remark that the spikes in the lower panels of Fig.~\ref{fig:GR_effects_ratio} appear because the RSD quadrupole crosses  zero, see Fig.~\ref{fig:RSD_pk}.
The larger effect on the quadrupole can be understood by noticing that lensing magnification contains a transverse Laplacian, which depends strongly on $\mu$ and it is therefore more sensitive to the non uniform weighting of higher multipoles.
The information carried by the lensing magnification effect is thus spread over several multipoles. Note for instance in Fig~\ref{fig:GR_effects_NOIC} that the lensing quadrupole, red line in the lower panels, is usually larger than the lensing monopole, red lines in the upper panels. Interestingly this could open the possibility to detect the lensing contribution by measuring the power spectrum at large scales for higher multipoles where the Newtonian effects are strongly suppressed.

Investigations with more realistic angular masks and selection function are certainly needed, but our analysis indicates that lensing magnification will likely not play a major role in cosmological parameter inference in spectroscopic surveys.

\subsection{The observed galaxy power spectrum}
\begin{figure}[!ht]
\begin{center}
\includegraphics[width=1.09\textwidth]{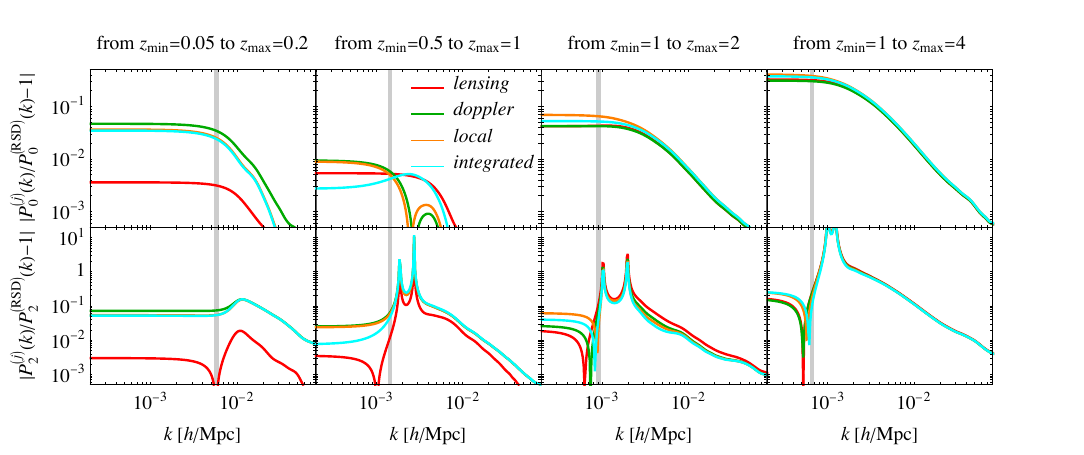}
\vspace{-0.75cm}
\caption{
We show the ratio between the multipoles (monopole on top panels and quadrupole in the bottom ones) sourced by RSD with respect to the other relativistic effects. We add the effect one by one, such that the lensing line contains also RSD, the doppler line also lensing and RSD, and so on. In the Integral Constraint we include the same effects.
The vertical lines correspond to the largest scales probed by the survey, \ie $2 \pi /\left( 2 r (z_{\rm max}) \sin\theta_{\rm max} \right)$.
}
\label{fig:GR_effects_ratio}
\end{center}
\end{figure}

In this section we finally discuss all the relativistic effects and their effect on the observed power spectrum.
As we can see from Figs.~\ref{fig:GR_effects_NOIC}~-~\ref{fig:GR_effects_ratio}, the relevance of the different effects changes with the survey volume and redshift coverage. In Fig.~\ref{fig:GR_effects_ratio}, as discussed in the previous Section, each line shows the ratio between the RSD prediction and RSD plus all the GR terms above in the Figure legend. For example the green line is the relative difference between the RSD term and RSD plus lensing and Doppler terms, and so on. 

At low redshifts the main contribution is due to the Doppler effects. This behaviour is well explained by the presence of the dependence on $1/(\HH r)$ of such terms. In our low redshift survey, the Doppler terms are approximately 5\% and 10\% of the RSD ones for the monopole and the quadrupole respectively.  They are therefore much larger than the lensing magnification. Indeed, lensing is an integrated effect from the source to the observer and its contribution growths by summing up several lensing deflections along the line of sight.
Local and integrated terms are subdominant at such low redshift.
Notice that all GR terms are proportional to $P(k) k^{n}$ for  some  $n<0$, until the window function dominates.

At intermediate redshifts, between $z_{\rm min}=0.5$ and $z_{\rm max}=1$, lensing become comparable to the Doppler terms, and the other local and integrated terms also grow. For this survey, GR corrections are still sub-\% for the monopole and approximately 10\% for the quadrupole on the largest scales. 
Interestingly, in our setup, the GR effects are minimized for a survey at intermediate redshifts, as the Doppler effect starts to decrease and lensing magnification has not yet taken over.

Moving to the high redshift Universe, in our survey between $z_{\rm min}=1$ and $z_{\rm max}=2$ or $z_{\rm min}=1$ and $z_{\rm max}=4$ the lensing magnification dominates over all other terms, and the conclusions we reached in the previous section apply. 

Our results indicate that for current and upcoming LSS surveys, cosmological inference from Baryon Acoustic Oscillations and the shape of the power spectrum on quasi-linear scales is largely unaffected by 
GR effects, which are sub per mille corrections to the RSD term.

\begin{figure}[!ht]
\begin{center}
  \includegraphics[width=1.09\textwidth]{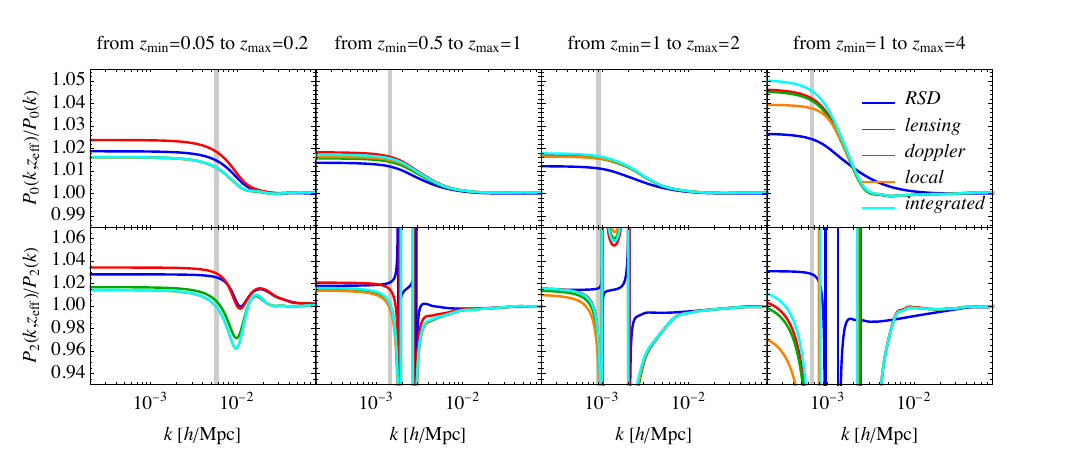}
  \vspace{-0.75cm}
\caption{
We show the accuracy of the effective redshift approximation (according to \eq{eq:def_zeff_th}) by including all the relativistic effects. Similarly to Fig.~\ref{fig:GR_effects_ratio} we add each effect one by one. The top panels refer to the monopole, while the bottom ones to the quadrupole. The vertical lines correspond to the largest scales probed by the survey, \ie $2 \pi /\left( 2 r (z_{\rm max}) \sin\theta_{\rm max} \right)$.
}
\label{fig:GR_zeff_effects_ratio}
\end{center}
\end{figure}

We conclude this section by investigating the accuracy of the effective redshift approximation once we include all the relativistic effects. Our findings are shown in Fig.~\ref{fig:GR_zeff_effects_ratio}, assuming the linear bias scales as $b_1(z) D(z) = 1.5$.
While we know that in principle the redshift evolution of the integrated terms, including lensing, is scale-dependent, because the latter are a relatively small correction on top of the standard RSD power spectrum the effective redshift approximation works extremely well also in the presence of relativistic corrections, with 5\% differences in the monopole only for the very extended survey between $z_{\rm min}=1$ and $z_{\rm max} =4$ shown in the rightmost column\footnote{We caution the reader that this approximation may fail if one is interested in the integrated terms only.}.

\section{Conclusions}
\label{sec:conclusions}
In this work we presented the first computation, in General Relativity, for the ensemble average of the Fourier space estimators traditionally used in the galaxy power spectrum analyses of spectroscopic surveys.
We clarified the relation between what is measured and what is predicted and how different estimators effectively perform different compressions of the data. 
Since GR corrections are important on large scales, a meaningful calculation must include observational effects like the presence of an angular mask, of a redshift selection function, and of the Integral Constraint. We implemented the presence of an arbitrary window function in the exact way, highlighting which assumptions could be employed to simplify the computation in certain situations.

A prerequisite for the GR calculation is the understanding of the IR divergences, and more generally of the IR sensitivity, of the three dimensional correlation function of the galaxy overdensity field. We showed for the first time that these effects are not present once all the terms in the expression for the galaxy number counts are taken into account, and the full calculation for any observable is finite and insensitive to the large scale cutoff of the theory.
We were able to trace back this cancellation  to the presence of the Weinberg adiabatic mode~\cite{Weinberg:2003sw} and of the Consistency relations for LSS~\cite{Creminelli2013,Peloso2013,Kehagias2013}. From these results, our calculations therefore inherit the underlying conditions of validity such as adiabatic and Gaussian initial conditions, and diffeomorphism invariance.
Our expression for the observed power spectrum differs from the one in Ref.~\cite{Grimm:2020ays}, with whom we therefore disagree on the interpretation of the relevance of the IR contributions to the power spectrum. Since our definition coincides with the power spectrum routinely measured in galaxy surveys we conclude that the power spectrum proposed in Ref.~\cite{Grimm:2020ays} does not necessarily correspond to an observable quantity. 

We then computed the observed power spectrum in four different hypothetical surveys, covering one third of the sky at different redshifts. We first confirmed earlier work in the literature on wide angle RSD, which we found to be a few \% of the total signal on very large scales. We also validated the series expansion approach to wide angle RSD, which is sub-\% accurate once the first few terms in the series have been included. We then checked the accuracy of using an effective redshift to model the clustering signal, and found that for our choice of radial selection functions it has a negligible impact, unless in a very extended survey between $z=1$ and $z=4$. The validity of the effective redshift approximation depends on both the redshift evolution of the sample and the selection function, therefore any conclusions on more realistic galaxy samples than what we considered here should rely on the full machinery provided in this work.

We then showed for the first time the impact of lensing magnification on the multipoles of the observed power spectrum, finding it is not a significant effect at low-$z$ but it can account for the 5$-$10\% of the monopole and the quadrupole at large scales for sufficiently realistic high redshift surveys. 

A summary of our results, including all GR effects, can be found in Fig.~\ref{fig:GR_effects_NOIC}~-~\ref{fig:GR_effects_ratio}. For low-redshift surveys the Doppler term is the most important GR contribution, up to 5$-$10\% of the total power, while at intermediate and high redshift lensing magnification becomes more significant. Other GR effects like ISW, time delays, and local terms are in general negligible.
Being the GR effects a small contribution to the observed multipoles, the effective redshift approximation remains valid in most cases considered. 

There are several directions along which this work can be extended. 
One of our main motivation was to understand the importance of GR effects when constraining local PNG with scale dependent bias. We presented the expression for the average of the Fourier Space estimator, but we cannot yet quantitatively address any possible degeneracy of GR terms with PNG because we miss an expression for the covariance of the measured power spectrum including GR effects and the window function. The mask indeed makes the covariance of the power spectrum non diagonal on very large scales, and so far only approximate analytical calculations in the flat sky limit have been proposed \cite{Wadekar:2019rdu}. Such an investigation is subject of work in progress by the authors~\cite{futureFNL}.

Our implementation of the Yamamoto estimator uses the endpoint LOS, but several other choices are possible. In particular, for analyses aiming at the Doppler terms a symmetric choice of LOS, like the bisector, is preferable because it sets to zero purely geometric dipoles. Also in this case the power spectrum can be measured with FFTs~\cite{Castorina2018,Philcox2021}, and all our calculations go through unchanged with the replacement of $s_1$ with the length of the new LOS. 
It is also well known that Yamamoto-like estimators are not statistically optimal on large scales~\cite{Tegmark:1997yq}. It is therefore natural to ask how to incorporate GR effects in optimal quadratic estimators of the power spectrum, which are traditionally implemented using a  flat-sky RSD model~\cite{Philcox2020}.
As a final remark we note that the full GR calculation is computationally much slower than the Newtonian one. However given the smallness of many GR terms, one could imagine resorting to the flat-sky approximation in some cases and obtain order of magnitudes speed up of the numerical calculations \cite{Jelic-Cizmek:2020jsn}.
We plan to return to these interesting problems in forthcoming publications.

\acknowledgments
ED (No.~171494 and~171506) acknowledges financial support from the Swiss National Science Foundation during the first stages of this project.
We would like to thank the participants of the workshop "Relativistic Aspects of Large Scale Structure", where this work was first presented, for discussion.
We thank Matteo Foglieni for finding several typos in the equations presented in Appendices of the first version of this work.

\newpage
\appendix

\section{Azimuthally symmetric window function}
\label{sec:Azymuthal_window}

In this appendix we provide explicit expressions for the convolution with an azimuthally symmetric window function
\be \label{eq:window_def}
\phi \left( \vs_1 \right) = \phi \left( s_1 \right) W \left( \theta_1 \right) =\phi \left( s_1 \right) \Theta \left( \theta_{\rm max} - \theta_1 \right) \, .
\ee
For this survey geometry \eq{eq:Fell_def} reduces to
\be
F_{\ell_1}\left( s_1, s \right) = \sum_{\ell_2 \ell_3} \left(
\begin{array}{ccc}
\ell_1 & \ell_2 & \ell_3 \\
0 &0&0
\end{array}
\right)^2 \left( 2 \ell_3 +1 \right) \phi_{\ell_2} \left( s_1 , s \right) A_{\ell_3} \left( \frac{s}{s_1} \right)
\ee
where
\be
A_\ell \left( x \right) = 2 \pi \int_{0}^{\theta_{\rm max}}\dd \theta_1 \sin \theta_1 \int_0^\pi\dd \theta \sin \theta \int_0^{2 \pi}\dd \phi \Theta \left( \cos \theta_2 - \cos \theta_{\rm max} \right) \mathcal{L}_\ell \left( \hat \vs_1 \cdot \hat \vs \right) 
\ee
and
\be
\phi_\ell \left( s_1 , s \right) = \frac{2 \ell +1 }{2} \int_{-1}^1\dd \mu \phi \left( s_2 \right) \mathcal{L}_\ell \left( \mu \right) \,.
\ee
For a top-hat redshift selection function we also have that
\be
\phi \left( s_1 \right) = \Theta \left( s_1 -s_{\rm min} \right) \Theta \left(s_{\rm max} - s_{1} \right) \, ,
\ee
and we can perform the integral analytically obtaining 
\bea
\phi \left( s_2 \right) =  \phi \left(  \sqrt{s_1^2 + s^2 + 2 s_1 s \mu } \right)= \sum_\ell \phi_\ell \left( s_1, s \right) \mathcal{L}_\ell \left( \mu \right) 
\eea
where 
\bea
\phi_\ell \left( s_1 , s \right) 
&=&\frac{2 \ell +1 }{2}
\Theta \left( s_{\rm max} - \left( s - s_1 \right) \right)
\nonumber \\
&&
\left[ 
\Theta \left( s_{\rm max} - s_2\left( s_1, s , 1 \right) \right) \Theta \left( s_2\left( s_1, s , -1\right)- s_{\rm min}  \right) \int_{-1}^1\dd \mu \ \mathcal{L}_\ell \left( \mu \right)
\right. \nonumber \\
&&
\left.
+
\Theta \left( -s_{\rm max} + s_2\left( s_1, s ,1 \right) \right) \Theta \left( s_2\left( s_1, s , -1\right)- s_{\rm min}  \right) \int_{-1}^{\left( s_{\rm max}^2 - s^2 -s_1^2\right) /\left( 2 s s_1\right) }\!\!\!\!\!\!\!\!\!  \!\!\!\!\!\!\!\!\!\dd \mu \ \mathcal{L}_\ell \left( \mu \right)
\right. \nonumber \\
&&
\left.
+
\Theta \left( s_{\rm max} - s_2\left( s_1, s ,1 \right) \right) \Theta \left(- s_2\left( s_1, s , -1\right)+ s_{\rm min}  \right) \int_{\left( s_{\rm min}^2 - s^2 -s_1^2\right) /\left( 2 s s_1\right)}^{1 }\!\!\!\!\!\!\!\!\!  \!\!\!\!\!\!\!\!\!\dd \mu \ \mathcal{L}_\ell \left( \mu \right)
\right. \nonumber \\
&&
\left.
+
\Theta \left( -s_{\rm max} + s_2\left( s_1, s , 1\right) \right) \Theta \left(- s_2\left( s_1, s ,-1 \right)+ s_{\rm min}  \right) \int_{\left( s_{\rm min}^2 - s^2 -s_1^2\right) /\left( 2 s s_1\right)}^{\left( s_{\rm max}^2 - s^2 -s_1^2\right) /\left( 2 s s_1\right) } \!\!\!\!\!\!\!\!\!   \!\!\!\!\!\!\!\!\!\dd \mu \ \mathcal{L}_\ell \left( \mu \right)
\right]
\nonumber \\
\eea

For this symmetric window function we can also pursuit a different path, without relying on the calculation of the multipoles of the correlation function. In the case of relativistic effects integrated from the source to the observer, such approach can lead to a simpler numerical evaluation.
In this case we start considering the expectation value of the power spectrum estimator
\bea
\langle \hat P_L \left( k \right) \rangle &=&
\frac{2 L +1}{ A} \int \frac{d \Omega_k}{4 \pi}\int\dd^3 s_1 \dd^3 s_2 \xi \left( s_1 , s , \hat \vs_1 \cdot  \hat \vs \right)   \phi \left( \vs_1 \right) \phi \left( \vs_2 \right) e^{-i \bk \cdot \vs }
\mathcal{L}_L \left( \hat \bk \cdot \hat \vs_1 \right)  \nonumber \\
&=&
\frac{2L+1}{A} \left( -i \right) ^L \int\dd^3 s_1 \dd^3s \xi \left( s_1,s, \hat \vs_1 \cdot \hat \vs \right) \phi \left( \vs_1 \right) \phi \left( \vs_2 \right) j_L \left( k s \right) \mathcal{L}_L \left( \hat \vs \cdot \hat \vs_1 \right) 
\eea
where in the second line we have changed the integration variables ($\dd^3 s_1  \dd^3s_2 \rightarrow \dd^3 s_1  \dd^3s  $).
By explicitly using the symmetry of the survey geometry we obtain 
\bea
\label{eq:estimator_new} 
\langle \hat P_L \left( k \right) \rangle &=&\frac{2L+1}{A} \left( -i \right) ^L 2 \pi  \int\dd s_1 s_1^2 \dd s s^2\dd\theta \sin\theta \dd \theta_1 \sin\theta_1\dd\varphi \xi \left( s_1,s, \hat \vs_1 \cdot \hat \vs \right)
\nonumber \\
&&
\phi \left( s_1 \right) \phi \left( s_2 \right) j_L \left( k s \right) 
W \left( \theta_1 \right) W \left( \theta_2 \right) 
\mathcal{L}_L \left( \hat \vs \cdot \hat \vs_1 \right) 
\nonumber \\
&=&
\frac{2L+1}{A} \left( -i \right) ^L 4 \pi  \int\dd s_1 s_1^2 \dd s s^2\dd \theta \sin\theta \dd \theta_1 \sin\theta_1 \frac{d y}{\sqrt{1-y^2}} \xi \left( s_1,s, \hat \vs_1 \cdot \hat \vs \right) 
\nonumber \\
&&
\phi \left( s_1 \right) \phi \left( s_2 \right) j_L \left( k s \right) 
W \left( \theta_1 \right) W \left( \theta_2 \right) 
\mathcal{L}_L \left( \hat \vs \cdot \hat \vs_1 \right) 
\nonumber \\
&=&
\frac{2L+1}{A} \left( -i \right) ^L 4 \pi  \int\dd s_1 s_1^2 \dd s s^2\dd \mu \xi \left( s_1,s, \mu  \right) \phi \left( s_1 \right) j_L \left( k s \right)  \phi \left( s_2 \right)\mathcal{L}_L \left(\mu  \right) 
\nonumber \\
&&
\int\dd \theta_1\dd \theta
W \left( \theta_1 \right) W \left( \theta_2 \right) \Theta \left( \mu - \cos \left( \theta + \theta_1 \right) \right) \Theta \left(  \cos \left( \theta - \theta_1 \right) -\mu \right)
\nonumber \\
&&
 \frac{\sin \theta_1 \sin \theta}{\sqrt{\left( \sin \theta \sin \theta_1 \right)^2 - \left( \cos \theta \cos \theta_1 - \mu \right)^2 }}
 \nonumber \\
&=&
\frac{2L+1}{A} \left( -i \right) ^L   \int\dd s_1 s_1^2 \dd s s^2\dd \mu \xi \left( s_1,s, \mu  \right) 
\nonumber \\
&&
\phi \left( s_1 \right) j_L \left( k s \right)  \phi \left( s_2 \right)\mathcal{L}_L \left(\mu  \right) 
F \left(\frac{s}{s_1} , \mu \right) 
\eea
where we have introduced 
\bea
F \left(x, \mu \right) &=& 4 \pi \int_0^{\theta_{\rm max}}\dd \theta_1 \int_0^{\pi}\dd \theta \ \Theta \left(  \frac{x \cos \theta + \cos \theta_1}{ \sqrt{x^2+ 1+ 2 x \mu }} - \cos \theta_{\rm max}\right) 
 \Theta \left( \mu - \cos \left( \theta + \theta_1 \right) \right)
  \nonumber \\
&&
\Theta \left(  \cos \left( \theta - \theta_1 \right) -\mu \right) \frac{\sin \theta_1 \sin \theta}{\sqrt{\left( \sin \theta \sin \theta_1 \right)^2 - \left( \cos \theta \cos \theta_1 - \mu \right)^2 }} \, .
\eea
We can see that this simplified approach fully agree with Eqs.~(\ref{eq:estimator_standard}-\ref{eq:Fell_def}).
We can easily relate \eq{eq:estimator_new} with \eq{eq:estimator_standard} through
\bea
 \phi \left( s_2 \right) \mathcal{L}_L \left( \mu \right) F \left( \frac{s}{s_1},\mu \right)  = \sum_{\ell, \ell_1} 
\left(
\begin{array}{ccc}
L & \ell & \ell_1 \\
0 &0&0
\end{array}
\right)^2 \left( 2 \ell_1 +1 \right) \frac{2 \ell +1}{2} \mathcal{L}_\ell\left( \mu \right)  F_{\ell_1} \left( s_1 , s \right) \, .
\eea
In particular, for the monopole ($L=0$) we have
\be
\label{eq:comparison_window}
F \left( \frac{s}{s_1}, \mu \right) \phi \left( s_2 \right)= \sum_\ell \frac{2 \ell +1}{2  } \mathcal{L}_\ell\left( \mu \right)  F_{\ell} \left( s_1 , s \right) \, .
\ee

%%%%%%%%%%%%%%%%%%%%%%

\newpage
{
\section{Relativistic correlation function}
\label{sec:correlation_function_appendix}
The full relativistic correlation function can be written as
\bea
&& \hspace{-0.8cm}\xi^{\mathcal{O}\mathcal{O}'} \left( s_1 , s_2 , \cos \theta = \hat \vs_1 \cdot \hat \vs_2 \right)
\nonumber \\
&=& \sum_{i,n} J_{ni}^{\mathcal{O}\mathcal{O}'} \left(s_1 , s_2 , \cos \theta \right) D_1 \left( s_1 \right) D_1 \left( s_2 \right) I^n_i \left( s \right) 
\nonumber \\
&&+
\sum_{i,n} J_{ni}^{\mathcal{O}\mathcal{O}'} \left(s_1 ,s_2,{ \cos\theta} \right) D_1 \left( s_1 \right)  I^n_i \left( s_1 \right) +
\sum_{i,n} J_{ni}^{\mathcal{O}\mathcal{O}'} \left(s_2 ,s_1 ,{ \cos\theta} \right) D_1 \left( s_2 \right)  I^n_i \left( s_2 \right) 
\nonumber \\
&&+
 D_1 \left( s_2 \right) \int_0^{s_1}\dd \chi_1 \sum_{i,n} J_{ni}^{\mathcal{O}\mathcal{O}'} \left(s_1 , s_2 , \cos \theta,\chi_1 \right) I^n_i \left( \Delta \chi_1  \right) 
\nonumber \\
&&+
 D_1 \left( s_1 \right)  \int_0^{s_2}\dd \chi_2 \sum_{i,n} J_{ni}^{\mathcal{O}\mathcal{O}'} \left(s_1 , s_2  , \cos \theta,\chi_2 \right) I^n_i \left( \Delta \chi_2 \right) 
 \nonumber \\
&&+
 \int_0^{s_1}\dd \chi_1 \sum_{i,n} J_{ni}^{\mathcal{O}\mathcal{O}'} \left(s_1 , \cos \theta,\chi_1 \right) I^n_i \left( \chi_1  \right) 
+
  \int_0^{s_2}\dd \chi_2 \sum_{i,n} J_{ni}^{\mathcal{O}\mathcal{O}'} \left(  s_2  , \cos \theta,\chi_2 \right) I^n_i \left( \chi_2 \right) 
\nonumber \\
&&
+
\int_0^{s_1}\dd \chi_1  \int_0^{s_2} \dd \chi_2 \sum_{i,n} J_{ni}^{\mathcal{O}\mathcal{O}'} \left(s_1 , s_2 , \cos \theta,\chi_1,\chi_2 \right)  I^n_i \left( \chi \right) 
\nonumber \\
&&+
J_{\sigma2}^{\mathcal{O}\mathcal{O}'}\left( s_1 , s_2, \cos\theta \right) \sigma_2   + J_{\sigma4}^{\mathcal{O}\mathcal{O}'} \left( s_1 , s_2, \cos\theta \right) \sigma_4 \, ,
\eea
where
\bea
\Delta \chi_2 &=& \sqrt{s_1^2+\chi_2^2 -2 s_1 \chi_2 \cos\theta}
\\
\Delta \chi_1  &=&\sqrt{\chi_1^2+s_2^2 -2 \chi_1 s_2 \cos\theta}
\\
\chi&=&\sqrt{\chi_1^2+\chi_2^2 -2 \chi_1 \chi_2 \cos\theta}
\eea
and we have introduced
\be
\sigma_i = \int \frac{\dd q }{2 \pi^2} q^{2-i} P \left( q \right)  \, .
\ee
The functions  $J_{ni}$ are defined in this Appendix.
In this section we summarize the relativistic contribution to the full-sky correlation function. The standard Newtonian contribution has been already derived in Section~\ref{subsec:densityRSD}.
To be consistent with previous derivations \cite{Tansella:2018sld}, we present them in terms of the system of coordinates $\left(s_1, s_2, \cos\theta= \hat \vs_1 \cdot \hat \vs_2 \right)$ \footnote{With respect to Ref.~\cite{Tansella:2018sld} we include also the terms evaluated at the observer positions. Beyond these additional terms, the reader may need to apply spherical Bessel recursion relations to compare our results with Ref.~\cite{Tansella:2018sld}.}. 

%%%%%%%%%%%%%%%%%%%%%%%
\subsection{Auto-correlation}
\subsubsection{Lensing}
\be
\xi^{\kappa \kappa}\left( s_1, s_2 , \cos \theta \right) =
\!\!\int_0^{r_1}\! \!d\chi_1\!
\int_0^{r_2} \!\!\dd \chi_2 
\left[
J^{\kappa \kappa}_{00} I^0_{0} \left( \Delta\chi \right) +J^{\kappa \kappa}_{02} I^0_{2} \left(\Delta\chi \right) +J^{\kappa \kappa}_{31} I^3_{1} \left( \Delta\chi \right) +J^{\kappa \kappa}_{22} I^{2}_2 \left( \Delta\chi \right) 
\right]
\ee
where $\Delta\chi=\sqrt{\chi_1^2 + \chi_2^2 - 2 \chi_1 \chi_2 \cos \theta}$ and (with $y=\cos\theta$)
\bea
J^{\kappa \kappa}_{00} &=& -\frac{3\chi_1^2\chi_2^2}{4 s_1 s_2 \Delta \chi ^4 a(\chi_1) a(\chi_2)} \left(y^2-1\right) \HH_0^4 \Omega_{M0}^2 D_1(\chi_1) (\chi_1-s_1) D_1(\chi_2) (\chi_2-s_2)
\nonumber \\
&&
(5 s_b(s_1)-2) (5 s_b(s_2)-2) \left(8 y \left(\chi_1^2+\chi_2^2\right)-9\chi_1\chi_2 y^2-7\chi_1\chi_2\right) \, ,
\\
%%%%%%%%%%%%%%
J^{\kappa \kappa}_{02} &=& -\frac{3\chi_1^2\chi_2^2 }{2 s_1 s_2 \Delta \chi ^4 a(\chi_1) a(\chi_2)}\left(y^2-1\right) \HH_0^4 \Omega_{M0}^2 D_1(\chi_1) (\chi_1-s_1) D_1(\chi_2) (\chi_2-s_2)
\nonumber \\
&&
(5 s_b(s_1)-2) (5 s_b(s_2)-2) \left(4 y \left(\chi_1^2+\chi_2^2\right)-3\chi_1\chi_2 y^2-5\chi_1\chi_2\right) \, ,
\\
%%%%%%%%%%%%%%
J^{\kappa \kappa}_{31} &=& \frac{9 y \Delta \chi ^2 \HH_0^4 \Omega_{M0}^2 D_1(\chi_1) (\chi_1-s_1) D_1(\chi_2) (\chi_2-s_2) (5 s_b(s_1)-2) (5 s_b(s_2)-2)}{s_1 s_2 a(\chi_1) a(\chi_2)} \, , \ \
\\
%%%%%%%%%%%%%%
J^{\kappa \kappa}_{22} &=& \frac{9\chi_1\chi_2 \HH_0^4 \Omega_{M0}^2 D_1(\chi_1)}{4 \Delta \chi ^4 a(\chi_1) a(\chi_2) s_1 s_2}  (\chi_1-s_1) D_1(\chi_2) (\chi_2-s_2) (5 s_b(s_1)-2) (5 s_b(s_2)-2)
\nonumber \\
&&
\left[2\chi_1^4 \left(7 y^2-3\right)-16\chi_1^3\chi_2 y \left(y^2+1\right)+\chi_1^2\chi_2^2 \left(11 y^4+14 y^2+23\right)
\right.
\nonumber \\
&& \left.
-16\chi_1\chi_2^3 y \left(y^2+1\right)+2\chi_2^4 \left(7 y^2-3\right)\right] \, .
\eea

%%%%%%%%%%%%%%%%%%%%%%%
\subsubsection{Doppler}
We consider now the contribution of the Doppler term (including the peculiar velocity of the observer).
By following the same approach we find
\bea
\xi^{\ndv \ndv}\left( s_1, s_2 , \cos \theta \right) &=& D_1 \left( s_1 \right)  D_1 \left( s_2 \right) \left[  J_{00}^{\ndv \ndv} \left( s_1, s_2 ,\cos\theta\right) I^0_{0} \left( s \right) 
+ J_{02}^{\ndv \ndv} \left( s_1, s_2 ,\cos\theta\right)I^0_{2} \left( s \right) \right.
\nonumber \\
&&
\left. \qquad
+J_{04}^{\ndv \ndv}\left( s_1, s_2 ,\cos\theta \right) I^0_{4} \left( s \right)
+J_{20}^{\ndv \ndv}\left( s_1, s_2 ,\cos\theta\right) I^2_{0} \left( s \right)
\right]
\nonumber \\
&&
+ D_1 \left( s_1 \right) \left[ J_{31}^{\ndv \ndv} \left( s_1 , s_2, \cos \theta \right) I^3_1 \left( s_1 \right)  
+ J_{11}^{\ndv \ndv} \left( s_1 , s_2,\cos \theta \right) I^1_1 \left( s_1 \right) 
\right.
\nonumber \\
&& 
\left. \qquad
+ J_{13}^{\ndv \ndv} \left( s_1 ,s_2, \cos \theta \right) I^1_3 \left( s_1 \right) 
\right]
\nonumber \\
&&
+D_1 \left( s_2 \right) \left[
 J_{31}^{\ndv \ndv} \left( s_2 ,s_1, \cos \theta \right) I^3_1 \left( s_2 \right)  
+ J_{11}^{\ndv \ndv} \left( s_2 ,s_1, \cos \theta \right) I^1_1 \left( s_2 \right) 
\right.
\nonumber \\
&& 
\left. \qquad
+ J_{13}^{\ndv \ndv} \left( s_2 ,s_1, \cos \theta \right) I^1_3 \left( s_2 \right) \right]
\nonumber \\
&& + J_{\sigma2}^{\ndv \ndv} \left( s_1 , s_2 , \cos \theta \right) \sigma_2 \, ,
\eea
where
\bea
J_{00}^{\ndv \ndv} \left( s_1 , s_2,y \right) &=&
\frac{1}{45}  f_1f_2\HH_1 \HH_2 \mathcal{R}_1 \mathcal{R}_2 \left(y^2 s_1 s_2-2 y \left(s_1^2+s_2^2\right)+3 s_1 s_2\right) \, ,
%%%%%%%%%%%%%%
\\
J_{02}^{\ndv \ndv} \left( s_1 , s_2,y \right)  &=&
\frac{2}{63}  f_1f_2\HH_1 \HH_2 \mathcal{R}_1 \mathcal{R}_2 \left(y^2 s_1 s_2-2 y \left(s_1^2+s_2^2\right)+3 s_1 s_2\right) \, ,
%%%%%%%%%%%%%%
\\
J_{04}^{\ndv \ndv} \left( s_1 , s_2,y \right)  &=&
\frac{1}{105}  f_1f_2\HH_1 \HH_2 \mathcal{R}_1 \mathcal{R}_2 \left(y^2 s_1 s_2-2 y \left(s_1^2+s_2^2\right)+3 s_1 s_2\right) \, ,
%%%%%%%%%%%%%%
\\
J_{20}^{\ndv \ndv}  \left( s_1 , s_2,y \right) &=&
\frac{1}{3} y s^2 f_1f_2\HH_1 \HH_2 \mathcal{R}_1 \mathcal{R}_2 \, ,
%%%%%%%%%%%%%%
\\
J_{31}^{\ndv \ndv} \left( s_1 , s_2,y \right)  &=&
-y f_0 \HH_0 s_1^2 f_1 \mathcal{H}_1 \mathcal{R}_1 (\mathcal{R}_2-5 s_b(s_2)+2) \, ,
%%%%%%%%%%%%%%
\\
J_{11}^{\ndv \ndv} \left( s_1 , s_2,y \right)  &=&
\frac{1}{5} y f_0 \HH_0 s_1^2 f_1\HH_1 \mathcal{R}_1 (\mathcal{R}_2-5 s_b(s_2)+2) \, ,
%%%%%%%%%%%%%%
\\
J_{13}^{\ndv \ndv} \left( s_1 , s_2,y \right)  &=&
\frac{1}{5} y f_0 \HH_0 s_1^2 f_1\HH_1 \mathcal{R}_1 (\mathcal{R}_2-5 s_b(s_2)+2) \, ,
%%%%%%%%%%%%%%
\\
J_{\sigma2}^{\ndv \ndv} \left( s_1 , s_2,y \right)  &=&
\frac{1}{3} y f_0^2 \HH_0^2 (\mathcal{R}_1-5 s_b(s_1)+2) (\mathcal{R}_2-5 s_b(s_2)+2) \, .
\eea

%%%%%%%%%%%%%%%%%%%%%%%
\subsubsection{Local Gravitational Potential}
The contribution of the local gravitational potential terms is determined by
\bea
\xi^{\phi \phi }\left( s_1, s_2 , \cos \theta \right) &=& D_1 \left( s_1 \right) D_1 \left( s_2 \right) J_{40}^{\phi \phi}\left( s_1, s_2 \right) I^4_{0} \left( s \right) 
\nonumber \\
&&
+  D_1 \left( s_2 \right)  J_{40}^{\phi_0 \phi}(s_1,s_2)  I^4_{0} \left( s_2 \right)
+  D_1 \left( s_1 \right)  J_{40}^{\phi_0 \phi}(s_2,s_1) I^4_{0} \left( s_1 \right)
\nonumber \\
&&
+ J_{\sigma4}^{\phi\phi} \left( s_1 , s_2 \right) \sigma_4  \, ,
\eea
where
\bea
J_{40}^{\phi \phi} \left( s_1 , s_2 \right)  &=& 
\frac{s^4}{4 a(s_1) a(s_2)}  
\nonumber \\
&&
\left(f_2 \left(2 a(s_2) (f_{\rm evo}(s_2)-3) \HH_2^2+3 \HH_0^2 \Omega_{M0}\right)+3 \HH_0^2 \Omega_{M0} (\mathcal{R}_2+5 s_b(s_2)-2)\right)
\nonumber \\
&&
\left(f_1 \left(2 a(s_1) (f_{\rm evo}(s_1)-3) \HH_1^2+3 \HH_0^2 \Omega_{M0}\right)+3 \HH_0^2 \Omega_{M0} (\mathcal{R}_1+5 s_b(s_1)-2)\right)  \, ,
\nonumber \\
&&
\\
%%%%%%%%%%%%%%%%%%%%
J_{40}^{\phi_0 \phi}\left( s_1 , s_2 \right) &=& 
\frac{\HH_0 s_2^4}{4 s_1 a(s_2)}  (\HH_0 s_1 (2 f_0-3 \Omega_{M0}) \mathcal{R}_1+2 f_0(5 s_b(s_1)-2)) 
\nonumber \\
&&
\left(f_2 \left(2 a(s_2) (f_{\rm evo}(s_2)-3) \HH_2^2+3 \HH_0^2 \Omega_{M0}\right)+3 \HH_0^2 \Omega_{M0} (\mathcal{R}_2+5 s_b(s_2)-2)\right) \, ,
\nonumber \\
&&
\\
%%%%%%%%%%%%%%%%%%%%
J_{\sigma 4}\left( s_1 , s_2 \right)&=& 
\frac{\HH_0^2 }{4 s_1 s_2}(\HH_0 s_1 (2 f_0-3 \Omega_{M0}) \mathcal{R}_1+2 f_0(5 s_b(s_1)-2))
\nonumber \\
&&
(\HH_0 s_2 (2 f_0-3 \Omega_{M0}) \mathcal{R}_2+2 f_0(5 s_b(s_2)-2)) \, .
\eea
%%%%%%%%%%%%%%%%%%%%%%%
\subsubsection{Integrated Gravitational Potential}
For the integrated terms we obtain
\be
\xi^{\int \!\phi \int \!\phi }\left( s_1, s_2 , \cos \theta \right) = \int_0^{s_1}\dd \chi_1  \int_0^{s_2}\dd \chi_2  J^{\int \!\phi \int \!\phi}_{40} \left( s_1,s_2,\chi_1,\chi_2 \right) I^{4}_0 \left( \Delta \chi \right) \, ,
\ee
where
\bea
J^{\int \!\phi \int \!\phi}_{40} \left( s_1,s_2,\chi_1,\chi_2 \right)&=&
\frac{9 \Delta \chi ^4 \HH_0^4 \Omega_{M0}^2 D_1(\chi_1) D_1(\chi_2)}{a(\chi_1) a(\chi_2) s_1 s_2}
(s_1 (f(\chi_1)-1) \HH(\chi_1) \mathcal{R}_1-5 s_b(s_1)+2)
\nonumber \\
&&
(s_2 (f(\chi_2)-1) \HH(\chi_2) \mathcal{R}_2-5 s_b(s_2)+2)\, .
\eea

%%%%%%%%%%%%%%%%%
\subsection{Cross-correlations}
Here we present the different cross-correlations. In the final result we need to account for the permutation on the two different source positions.

\subsubsection{Standard Newtonian x Lensing}

We consider the correlation between the standard Newtonian terms and the lensing magnification.
\be
\xi^{\delta \kappa} \left( s_1, s_2 , \cos \theta \right)
=
 D_1 \left( s_1 \right)  \int_0^{s_2}\dd \chi_2
 \left[ J^{\delta \kappa}_{00} I_{00}  \left( \Delta \chi_2 \right) 
 + J^{\delta \kappa}_{02} I^0_{2} \left( \Delta \chi_2 \right) +
  J^{\delta \kappa}_{04} I^0_{4}  \left( \Delta \chi_2 \right) 
 \right] \, ,
\ee
where
\bea
J^{\delta \kappa}_{00}\left( s_1 , s_2, y,\chi_2 \right) &=&
\frac{\HH_0 ^2 \Omega_{M0} D_1 (\chi_2)}{5 a(\chi_2 ) s_2} (\chi_2 -s_2 )    (5 s_b (s_2 )-2) 
\nonumber \\
&&
\left(f(s_1 ) \left(\chi_2  \left(3 y^2-1\right)-3 y s_1 \right)-5 y s_1  b_1 (s_1 )\right) \, ,
\\
%%%%%%%%%%%%%%%%%%%%%%%%
J^{\delta \kappa}_{02}\left( s_1 , s_2, y,\chi_2 \right) &=&
\frac{\HH_0 ^2 \Omega_{M0} D_1 (\chi_2 ) }{14 \Delta\chi_2 ^2 a(\chi_2 ) s_2 }(\chi_2 -s_2 )    (5 s_b (s_2 )-2) 
\nonumber \\
&&
\left(7 s_1  b_1 (s_1 ) \left(-2 \chi_2 ^2 y+\chi_2  \left(y^2+3\right) s_1 -2 y s_1 ^2\right)
\right.
\nonumber \\
&&
\left.
+f(s_1 ) \left(4 \chi_2 ^3 \left(3 y^2-1\right)-2 \chi_2 ^2 y \left(3 y^2+8\right) s_1 
\right. \right.
\nonumber \\
&&
\left. \left.
+\chi_2  \left(9 y^2+11\right) s_1 ^2-6 y s_1 ^3\right)\right) \, ,
\\
%%%%%%%%%%%%%%%%%%%%%%%%
J^{\delta \kappa}_{04}\left( s_1 , s_2, y,\chi_2 \right)  &=&
\frac{3 \HH_0 ^2 \Omega_{M0} D_1 (\chi_2 )}{70 \Delta\chi_2 ^4 a(\chi_2 )s_2 } (\chi_2 -s_2 )    f(s_1 ) (5 s_b (s_2 )-2) 
\nonumber \\
&&
\left(\chi_2 ^5 \left(6 y^2-2\right)+6 \chi_2 ^4 y \left(y^2-3\right) s_1 -\chi_2 ^3 \left(y^4+12 y^2-21\right) s_1 ^2
\right.
\nonumber \\
&&
\left.
+2 \chi_2 ^2 y \left(y^2+3\right) s_1 ^3-12 \chi_2  s_1 ^4+4 y s_1 ^5\right) \, .
\eea

\subsubsection{Standard Newtonian x Doppler}

\bea
\xi^{\delta \ndv}\left( s_1, s_2 , \cos \theta \right) &=& 
D_1 \left( s_1 \right) D_1 \left( s_2 \right) \left[ J^{\delta \ndv}_{00} I^0_{0} \left( s \right) + J^{\delta \ndv}_{02} I^0_2 \left( s \right) 
+J^{\delta \ndv}_{04} I^0_4 \left( s \right) \right]
\nonumber \\
&&
+ D_1 \left( s_1 \right) \left[ J^{\delta \ndv}_{11} I^1_{1} \left( s_1 \right) +  J^{\delta \ndv}_{13} I^1_{3} \left( s_1 \right) \right] \, ,
\eea
where
 \bea
J^{\delta \ndv}_{00} \left( s_1 , s_2 , y \right)  &=&
\frac{1}{15}         f_2  \HH_2  \mathcal{R}_2  \left(5 b_1 (s_1 ) (s_2 -y s_1 )+f_1  \left(2 y^2 s_2 -3 y s_1 +s_2 \right)\right) \, ,
\\
%%%%%%%%%%%%%%%%%%%%%
J^{\delta \ndv}_{02} \left( s_1 , s_2 , y \right)&=&
\frac{     f_2  \HH_2  \mathcal{R}_2}{21 s^2}  \left(7 b_1 (s_1 ) (y s_1 -s_2 ) \left(2 y s_1  s_2 -s_1 ^2-s_2 ^2\right)
\right.
\nonumber \\
&&
\left. 
+f_1  \left(\left(10 y^2-1\right) s_1 ^2 s_2 -y \left(5 y^2+4\right) s_1  s_2 ^2+\left(y^2+2\right) s_2 ^3-3 y s_1 ^3\right)\right) \, , \quad
\\
%%%%%%%%%%%%%%%%%%%%%
J^{\delta \ndv}_{04} \left( s_1 , s_2 , y \right)&=&
\frac{    f_1  f_2  \HH_2  \mathcal{R}_2 }{35 s^2} \left(-2 \left(y^2+2\right) s_1 ^2 s_2 +y \left(y^2+5\right) s_1  s_2 ^2
\right.
\nonumber \\
&&
\left. 
+\left(1-3 y^2\right) s_2 ^3+2 y s_1 ^3\right) \, ,
\\
%%%%%%%%%%%%%%%%%%%%%
J^{\delta \ndv}_{11} \left( s_1 , s_2 , y \right)&=&
\frac{1}{5} y f_0  \HH_0   s_1   (5 b_1 (s_1 )+3 f_1 ) (\mathcal{R}_2 -5 s_b (s_2 )+2) \, ,
\\
%%%%%%%%%%%%%%%%%%%%%
J^{\delta \ndv}_{13} \left( s_1 , s_2 , y \right) &=&
\frac{1}{5} (-2) y f_0  \HH_0   s_1  f_1  (\mathcal{R}_2 -5 s_b (s_2 )+2) \, .
\eea
}

\subsubsection{Standard Newtonian x Local Gravitational Potential}
\bea
\xi^{\delta \phi}\left( s_1, s_2 , \cos \theta \right) &=& 
D_1 \left( s_1 \right) D_1 \left( s_2 \right) \left[ J^{\delta \phi}_{00} I^0_{0} \left( s \right) + J^{\delta \phi}_{02} I^0_2 \left( s \right) 
+J^{\delta \phi}_{04} I^0_4 \left( s \right)
+J^{\delta \phi}_{20} I^2_0 \left( s \right)\right]
\nonumber \\
&&
+ D_1 \left( s_1 \right) \left[ J^{\delta \phi}_{31} I^3_{1} \left( s_1 \right)+J^{\delta \phi}_{11} I^1_{1} \left( s_1 \right) +  J^{\delta \phi}_{13} I^1_{3} \left( s_1 \right) \right] \, ,
\eea
where
 \bea
J^{\delta \phi}_{00} \left( s_1 , s_2 , y \right)  &=&
\frac{          f_1}{90 a(s_2)} \left(\left(3 y^2-1\right) s_2^2-4 y s_1 s_2+2 s_1^2\right)
\nonumber \\
&&
\left(f_2 \left(2 a(s_2) (f_{\rm evo}(s_2)-3) \HH_2^2+3 \HH_0^2 \Omega_{M0}\right)+3 \HH_0^2 \Omega_{M0} (\mathcal{R}_2+5 s_b(s_2)-2)\right) \, ,
\nonumber \\
&&
\\
%%%%%%%%%%%%%%%%%%%%%
J^{\delta \phi}_{02} \left( s_1 , s_2 , y \right)&=&
\frac{          f_1}{63 a(s_2)} \left(\left(3 y^2-1\right) s_2^2-4 y s_1 s_2+2 s_1^2\right)
\nonumber \\
&&
\left(f_2 \left(2 a(s_2) (f_{\rm evo}(s_2)-3) \HH_2^2+3 \HH_0^2 \Omega_{M0}\right)+3 \HH_0^2 \Omega_{M0} (\mathcal{R}_2+5 s_b(s_2)-2)\right) \, ,
\nonumber \\
&&
\\
%%%%%%%%%%%%%%%%%%%%%
J^{\delta \phi}_{04} \left( s_1 , s_2 , y \right)&=&
\frac{          f_1}{210 a(s_2)} \left(\left(3 y^2-1\right) s_2^2-4 y s_1 s_2+2 s_1^2\right)
\nonumber \\
&&
\left(f_2 \left(2 a(s_2) (f_{\rm evo}(s_2)-3) \HH_2^2+3 \HH_0^2 \Omega_{M0}\right)+3 \HH_0^2 \Omega_{M0} (\mathcal{R}_2+5 s_b(s_2)-2)\right) \, ,
\nonumber \\
&&
\\
%%%%%%%%%%%%%%%%%%%%%
J^{\delta \phi}_{20} \left( s_1 , s_2 , y \right)&=&
-\frac{ (3 b_1(s_1)+f_1) }{6 a(s_2)} \left(-2 y s_1 s_2+s_1^2+s_2^2\right)
\nonumber \\
&&
\left(f_2 \left(2 a(s_2) (f_{\rm evo}(s_2)-3) \HH_2^2+3 \HH_0^2 \Omega_{M0}\right)+3 \HH_0^2 \Omega_{M0} (\mathcal{R}_2+5 s_b(s_2)-2)\right) \, ,
\nonumber \\
&&
\\
%%%%%%%%%%%%%%%%%%%%%
J^{\delta \phi}_{31} \left( s_1 , s_2 , y \right)&=&
-\frac{\HH_0 s_1^2      (3 b_1(s_1)+f_1) (\HH_0 s_2 (2 f_0-3 \Omega_{M0}) \mathcal{R}_2+2 f_0 (5 s_b(s_2)-2))}{2 s_2} \, ,
\\
%%%%%%%%%%%%%%%%%%%%%
J^{\delta \phi}_{11} \left( s_1 , s_2 , y \right)&=&
\frac{\HH_0 s_1^2      (b_1(s_1)+f_1) (\HH_0 s_2 (2 f_0-3 \Omega_{M0}) \mathcal{R}_2+2 f_0 (5 s_b(s_2)-2))}{10 s_2} \, ,
\\
%%%%%%%%%%%%%%%%%%%%%
J^{\delta \phi}_{13} \left( s_1 , s_2 , y \right) &=&
\frac{\HH_0 s_1^2      (b_1(s_1)+f_1) (\HH_0 s_2 (2 f_0-3 \Omega_{M0}) \mathcal{R}_2+2 f_0 (5 s_b(s_2)-2))}{10 s_2} \, .
\eea

\subsubsection{Standard Newtonian x Integrated Gravitational Potential}
\bea
\xi^{\delta \int \! \phi}\left( s_1, s_2 , \cos \theta \right) &=&
D_1 \left( s_1 \right) \int_0^{s_2}\dd \chi_2 \left[ J^{\delta \int \! \phi}_{00} I^0_{0} \left( \Delta\chi_2 \right) + J^{\delta \int \! \phi}_{02} I^0_2 \left( \Delta\chi_2 \right) 
\right.
\nonumber \\
&& \qquad \qquad 
\left.
+J^{\delta \int \! \phi}_{04} I^0_4 \left( \Delta\chi_2 \right)
+J^{\delta \int \! \phi}_{20} I^2_0 \left( \Delta\chi_2 \right)\right] \, ,
\eea
where
\bea
J^{\delta \int \! \phi}_{00} \left( s_1 , s_2, y ,\chi_2 \right) &=&
\frac{\HH_0^2 \Omega_{M0} D_1(\chi_2)    f_1}{15 a(\chi_2) s_2} \left(\chi_2^2 \left(3 y^2-1\right)-4 \chi_2 y s_1+2 s_1^2\right)
\nonumber \\
&&
(s_2 (f(\chi_2)-1) \HH(\chi_2) \mathcal{R}_2-5 s_b(s_2)+2) \, ,
\\
%%%%%%%%%%%%%%%%%
J^{\delta \int \! \phi}_{02}\left( s_1 , s_2, y ,\chi_2 \right) &=&
\frac{2 \HH_0^2 \Omega_{M0} D_1(\chi_2)    f_1}{21 a(\chi_2)s_2} \left(\chi_2^2 \left(3 y^2-1\right)-4 \chi_2 y s_1+2 s_1^2\right)
\nonumber \\
&&
(s_2 (f(\chi_2)-1) \HH(\chi_2) \mathcal{R}_2-5 s_b(s_2)+2) \, ,
\\
%%%%%%%%%%%%%%%%%
J^{\delta \int \! \phi}_{04} \left( s_1 , s_2, y ,\chi_2 \right)&=&
\frac{\HH_0^2 \Omega_{M0} D_1(\chi_2)    f_1}{35 a(\chi_2)s_2} \left(\chi_2^2 \left(3 y^2-1\right)-4 \chi_2 y s_1+2 s_1^2\right)
\nonumber \\
&&
(s_2 (f(\chi_2)-1) \HH(\chi_2) \mathcal{R}_2-5 s_b(s_2)+2) \, ,
\\
%%%%%%%%%%%%%%%%%
J^{\delta \int \! \phi}_{20}\left( s_1 , s_2, y ,\chi_2 \right) &=&
-\frac{\Delta\chi_2^2 \HH_0^2 \Omega_{M0} D_1(\chi_2) }{a(\chi_2)s_2}   (3 b_1(s_1)+f_1) 
\nonumber \\
&&
(s_2 (f(\chi_2)-1) \HH(\chi_2) \mathcal{R}_2-5 s_b(s_2)+2) \, .
%%%%%%%%%%%%%%%%%
\eea

\subsubsection{Lensing x Doppler}
\bea
\xi^{\kappa \ndv} \left( s_1 , s_2, \cos \theta \right) &=& D_1 \left( s_2 \right) \int_0^{s_1}\dd \chi_1 
\left[
J^{\kappa \ndv}_{00} I^0_0 \left( \Delta \chi_1 \right) +
J^{\kappa \ndv}_{02} I^0_2 \left( \Delta \chi_1 \right) 
\right.
\nonumber \\
&&
\left.
+
J^{\kappa \ndv}_{04} I^0_4 \left( \Delta \chi_1 \right) 
+
J^{\kappa \ndv}_{20} I^2_0 \left( \Delta \chi_1 \right) 
\right]
%\nonumber \\
%&& 
+ \int_0^{s_1}\dd \chi_1 J^{\kappa \ndv}_{31} I^3_1 \left( \chi_1 \right) \, ,
\eea
where 
\bea
J^{\kappa \ndv}_{00} \left( s_1 , s_2 , y, \chi_1\right)  &=&
\frac{\HH_0^2 \Omega_{M0} D_1(\chi_1)}{15 a(\chi_1) s_1 } (\chi_1\!- \! s_1)    f_2 \HH_2 \mathcal{R}_2 (5 s_b(s_1)-2) \nonumber \\
&&
\left(\chi_1^2   y+\chi_1 \left(4   y^2-3\right) s_2-2   y s_2^2\right) \, ,
\\
%%%%%%%%%%%%%%%%%
J^{\kappa \ndv}_{02}
\left( s_1 , s_2 , y, \chi_1\right) &=&
\frac{\HH_0^2 \Omega_{M0} D_1(\chi_1)}{42 \Delta\chi_1^2 a(\chi_1) s_1} (\chi_1\!- \! s_1)    f_2 \HH_2 \mathcal{R}_2 (5 s_b(s_1)-2) 
\nonumber \\
&&
\left(4 \chi_1^4   y+4 \chi_1^3 \left(2   y^2-3\right) s_2+\chi_1^2   y \left(11-23   y^2\right) s_2^2
\right.
\nonumber \\
&&
\left.
+\chi_1 \left(23   y^2-3\right) s_2^3-8   y s_2^4\right) \, ,
\\
%%%%%%%%%%%%%%%%%
J^{\kappa \ndv}_{04}
\left( s_1 , s_2 , y, \chi_1\right) &=&
\frac{\HH_0^2 \Omega_{M0} D_1(\chi_1) }{70 \Delta\chi_1^2 a(\chi_1)s_1}(\chi_1\!- \! s_1)    f_2 \HH_2 \mathcal{R}_2 (5 s_b(s_1)-2) 
\nonumber \\
&&
\left(2 \chi_1^4   y+2 \chi_1^3 \left(2   y^2-3\right) s_2-\chi_1^2   y \left(  y^2+5\right) s_2^2
\right.
\nonumber \\
&& \left.
+\chi_1 \left(  y^2+9\right) s_2^3-4   y s_2^4\right) \, ,
\\
%%%%%%%%%%%%%%%%%
J^{\kappa \ndv}_{20}
\left( s_1 , s_2 , y, \chi_1\right) &=&
\frac{  y \Delta\chi_1^2 \HH_0^2 \Omega_{M0} D_1(\chi_1)}{a(\chi_1)s_1} (\chi_1\!- \! s_1)    f_2 \HH_2 \mathcal{R}_2 (5 s_b(s_1)-2) \, ,
\\
%%%%%%%%%%%%%%%%%
J^{\kappa \ndv}_{31} \left( s_1 , s_2 , y, \chi_1\right) &=&
-\frac{3 \chi_1^2   y f_0 \HH_0^3 \Omega_{M0} D_1(\chi_1) }{a(\chi_1)s_1}(\chi_1-s_1) (5 s_b(s_1)-2)
\nonumber \\
&&
(\mathcal{R}_2-5 s_b(s_2)+2) \, .
\eea
\subsubsection{Lensing x Local Gravitational Potential}
\bea
\xi^{\kappa \phi} \left( s_1 , s_2, \cos \theta \right) &=& D_1 \left( s_2 \right) \int_0^{s_1}\dd \chi_1 \left[ 
J^{\kappa \phi}_{00} I^0_0 \left( \Delta \chi_1 \right) +
J^{\kappa \phi}_{02} I^0_2 \left( \Delta \chi_1 \right) 
\right.
\nonumber \\
&&
\left.
\qquad
+
J^{\kappa \phi}_{04} I^0_4 \left( \Delta \chi_1 \right) 
+
J^{\kappa \phi}_{20} I^2_0 \left( \Delta \chi_1 \right) 
\right] \, ,
\eea
where
\bea
J^{\kappa \phi}_{00} \left( s_1,s_2, y, \chi_1\right) &=& \frac{\HH_0^2 \Omega_{M0} s_2 D_1(\chi_1) }{60 a(\chi_1) a(s_2) s_1} (\chi_1\!- \! s_1)     (5 s_b(s_1)-2) \left(2 \chi_1^2 y-\chi_1 \left(y^2+3\right) s_2+2 y s_2^2\right) 
\nonumber \\
&&
\left(f_2\left(2 a(s_2) (f_{\rm evo}(s_2)-3) \HH_2^2+3 \HH_0^2 \Omega_{M0}\right)
\right.
\nonumber \\
&&
\left.
+3 \HH_0^2 \Omega_{M0} (\mathcal{R}_2+5 s_b(s_2)-2)\right)  \, ,
\\
%%%%%%%%%%%%%%%%%%%
J^{\kappa \phi}_{02} \left( s_1,s_2 ,y, \chi_1\right) &=& \frac{\HH_0^2 \Omega_{M0} s_2 D_1(\chi_1) }{42 a(\chi_1) a(s_2)s_1} (\chi_1\!- \! s_1)     (5 s_b(s_1)-2) \left(2 \chi_1^2 y-\chi_1 \left(y^2+3\right) s_2+2 y s_2^2\right)
\nonumber \\
&&
\left(f_2\left(2 a(s_2) (f_{\rm evo}(s_2)-3) \HH_2^2+3 \HH_0^2 \Omega_{M0}\right)
\right.
\nonumber \\
&&
\left.
+3 \HH_0^2 \Omega_{M0} (\mathcal{R}_2+5 s_b(s_2)-2)\right) \, ,
\\
%%%%%%%%%%%%%%%%%%%
J^{\kappa \phi}_{04} \left( s_1,s_2 ,y, \chi_1\right) &=& \frac{\HH_0^2 \Omega_{M0} s_2 D_1(\chi_1)}{140 a(\chi_1) a(s_2)s_1} (\chi_1\!- \! s_1)     (5 s_b(s_1)-2) \left(2 \chi_1^2 y-\chi_1 \left(y^2+3\right) s_2+2 y s_2^2\right) 
\nonumber \\
&&
\left(f_2\left(2 a(s_2) (f_{\rm evo}(s_2)-3) \HH_2^2
+3 \HH_0^2 \Omega_{M0}\right)
\right.
\nonumber \\
&&
\left.
+3 \HH_0^2 \Omega_{M0} (\mathcal{R}_2+5 s_b(s_2)-2)\right) \, ,
\\
%%%%%%%%%%%%%%%%%%%
J^{\kappa \phi}_{20} \left( s_1,s_2 ,y, \chi_1\right) &=& \frac{y \Delta \chi_1^2 \HH_0^2 \Omega_{M0} s_2 D_1(\chi_1)}{2 a(\chi_1) a(s_2)s_1} (\chi_1-s_1)  (5 s_b(s_1)-2) 
\nonumber \\ 
&&
\left(f(s_2) \left(2 a(s_2) (f_{\rm evo}(s_2)-3) \HH_2^2 +3 \HH_0^2 \Omega_{M0}\right)
\right.
\nonumber \\ 
&& \left.
+3 \HH_0^2 \Omega_{M0} (\mathcal{R}_2+5 s_b(s_2)-2)\right) \, .
%%%%%%%%%%%%%%%%%%%
\eea

\subsubsection{Lensing x Integrated Gravitational Potential}
\be
\xi^{\kappa \int \! \phi} \left( s_1 , s_2 , \cos \theta \right) = \int_0^{s_1}\dd \chi_1 \int_0^{s_2}\dd \chi_2 \left[ J_{31}^{\kappa \int \! \phi} I^3_1 \left( \Delta \chi \right) 
+
 J_{22}^{\kappa \int \! \phi} I^2_2 \left( \Delta \chi \right) \right] \, ,
\ee
where 
\bea
J_{31}^{\kappa \int \! \phi} \left( s_1 , s_2 , \cos\theta, \chi_1 , \chi_2 \right)  &=&\frac{9 \chi_2 y \Delta \chi ^2 \HH_0^4 \Omega_{M0}^2 }{a(\chi_1) a(\chi_2)s_1 s_2} D_1(\chi_1) (\chi_1-s_1) D_1(\chi_2) (5 s_b(s_1)-2)
\nonumber \\
&&
(s_1 (f(\chi_1)-1) \mathcal{H}(\chi_1) \mathcal{R}_-5 s_b(s_1)+2) \, ,
\\
%%%%%%%%%%%%%
J_{22}^{\kappa \int \! \phi} \left( s_1 , s_2 , \cos\theta, \chi_1 , \chi_2 \right)  &=&
\frac{9 \chi_1 \chi_2^2 \left(y^2-1\right) \HH_0^4 \Omega_{M0}^2}{2 a(\chi_1) a(\chi_2) s_1 s_2} D_1(\chi_1) (\chi_1-s_1) D_1(\chi_2) (5 s_b(s_1)-2) 
\nonumber \\
&&
(s_2 (f(\chi_2)-1) \HH(\chi_2) \mathcal{R}_2-5 s_b(s_2)+2) \, .
\eea

\subsubsection{Doppler x Local Gravitational Potential}
\bea
\xi^{\ndv \phi} \left( s_1 , s_2 , \cos\theta \right) &=& D_1 \left( s_1 \right) D_1 \left( s_2 \right) \left[ 
J^{\ndv \phi}_{00} I^0_0 \left( s \right) 
+ J^{\ndv \phi}_{02} I^0_2 \left( s \right) 
+ J^{\ndv \phi}_{04} I^0_4 \left( s \right) 
+ J^{\ndv \phi}_{20} I^2_0 \left( s \right) 
\right]
\nonumber \\
&& +
D_1 \left( s_1 \right) J^{\ndv \phi_0 }_{31} I^3_1 \left( s_1 \right) 
+ D_1 \left( s_2 \right) J^{\ndv_0 \phi }_{31} I^3_1 \left( s_2 \right) \, ,
\eea
where
\bea
J^{\ndv \phi}_{00} \left( s_1 , s_2 , y \right) &=&
-\frac{s^2         f_1 \HH_1 \mathcal{R}_1}{90 a(s_2)} (s_1-y s_2) \left(f_2 \left(2 a(s_2) (f_{\rm evo}(s_2)-3) \HH_2^2+3 \HH_0^2 \Omega_{M0}\right)
\right.
\nonumber \\
&& \left.
+3 \HH_0^2 \Omega_{M0} (\mathcal{R}_2+5 s_b(s_2)-2)\right) \, ,
%%%%%%%%%%%%%%%%%%%%
\\
J^{\ndv \phi}_{02} \left( s_1 , s_2 , y \right) &=&
-\frac{s^2         f_1 \HH_1 \mathcal{R}_1 }{63 a(s_2)} (s_1-y s_2) \left(f_2 \left(2 a(s_2) (f_{\rm evo}(s_2)-3) \HH_2^2+3 \HH_0^2 \Omega_{M0}\right)
\right.
\nonumber \\
&& \left. 
+3 \HH_0^2 \Omega_{M0} (\mathcal{R}_2+5 s_b(s_2)-2)\right) \, ,
%%%%%%%%%%%%%%%%%%%%
\\
J^{\ndv \phi}_{04} \left( s_1 , s_2 , y \right) &=&
-\frac{s^2         f_1 \HH_1 \mathcal{R}_1}{210 a(s_2)} (s_1-y s_2) \left(f_2 \left(2 a(s_2) (f_{\rm evo}(s_2)-3) \HH_2^2+3 \HH_0^2 \Omega_{M0}\right)
\right.
\nonumber \\
&& \left. 
+3 \HH_0^2 \Omega_{M0} (\mathcal{R}_2+5 s_b(s_2)-2)\right) \, ,
%%%%%%%%%%%%%%%%%%%%
\\
J^{\ndv \phi}_{20} \left( s_1 , s_2 , y \right) &=&
-\frac{s^2         f_1 \HH_1 \mathcal{R}_1}{6 a(s_2)} (s_1-y s_2) \left(f_2 \left(2 a(s_2) (f_{\rm evo}(s_2)-3) \HH_2^2+3 \HH_0^2 \Omega_{M0}\right)
\right.
\nonumber \\
&& \left. 
+3 \HH_0^2 \Omega_{M0} (\mathcal{R}_2+5 s_b(s_2)-2)\right) \, ,
%%%%%%%%%%%%%%%%%%%%
\\
J^{\ndv \phi_0 }_{31}\left( s_1 , s_2 , y \right)  &=&
\frac{\HH_0 s_1^3     f_1 \HH_1 \mathcal{R}_1 (\HH_0 s_2 (3 \Omega_{M0}-2 f_0) \mathcal{R}_2+2 f_0 (2-5 s_b(s_2)))}{2 s_2} \, ,
%%%%%%%%%%%%%%%%%%%%
\\
J^{\ndv_0 \phi }_{31} \left( s_1 , s_2 , y \right)  &=&
-\frac{y f_0 \HH_0 s_2^3   }{2 a(s_2)}  (\mathcal{R}_1-5 s_b(s_1)+2) \left(f_2 \left(2 a(s_2) (f_{\rm evo}(s_2)-3) \HH_2^2+3 \HH_0^2 \Omega_{M0}\right)
\right.
\nonumber \\
&& \left. 
+3 \HH_0^2 \Omega_{M0} (\mathcal{R}_2+5 s_b(s_2)-2)\right) \, .
\eea

\subsubsection{Doppler x Integrated Gravitational Potential}

\bea
\xi^{\ndv \int \! \phi} \left( s_1 , s_2, \cos\theta \right) &=& D_1 \left( s_1 \right) \int_0^{s_2}\dd \chi_2\left[
J^{\ndv \int \! \phi}_{00} I^0_0 \left( \Delta \chi_2 \right) 
+J^{\ndv \int \! \phi}_{02} I^0_2 \left( \Delta \chi_2 \right) 
\right. 
\nonumber \\
&&
\left.
+J^{\ndv \int \! \phi}_{04} I^0_4 \left( \Delta \chi_2 \right) 
+J^{\ndv \int \! \phi}_{20} I^2_0 \left( \Delta \chi_2 \right) 
\right]
+ \int_0^{s_2}\dd \chi_2  J^{\ndv \int \! \phi}_{31} I^3_1 \left(  \chi_2 \right)  \, ,
\nonumber \\
&&
\eea
where
\bea
J^{\ndv \int \! \phi}_{00} \left( s_1 , s_2 , y ,\chi_2 \right) &=&
\frac{\Delta\chi_2^2 \HH_0^2 \Omega_{M0} D_1(\chi_2)     f_1 \HH_1 \mathcal{R}_1}{15 s_2 a(\chi_2)} (\chi_2 y-s_1)
\nonumber \\
&&
(s_2 (f(\chi_2)-1) \HH(\chi_2) \mathcal{R}_2-5 s_b(s_2)+2) \, ,
%%%%%%%%%%%%%%%%%%%%%
\\
J^{\ndv \int \! \phi}_{02} \left( s_1 , s_2 , y ,\chi_2 \right) &=&
\frac{2 \Delta\chi_2^2 \HH_0^2 \Omega_{M0} D_1(\chi_2)     f_1 \HH_1 \mathcal{R}_1 }{21 s_2 a(\chi_2)}(\chi_2 y-s_1)
\nonumber \\
&&
(s_2 (f(\chi_2)-1) \HH(\chi_2) \mathcal{R}_2-5 s_b(s_2)+2) \, ,
\\
%%%%%%%%%%%%%%%%%%%%%
J^{\ndv \int \! \phi}_{04} \left( s_1 , s_2 , y ,\chi_2 \right) &=&
\frac{\Delta\chi_2^2 \HH_0^2 \Omega_{M0} D_1(\chi_2)     f_1 \HH_1 \mathcal{R}_1 }{35 s_2 a(\chi_2)}
\nonumber \\
&&
(\chi_2 y-s_1) (s_2 (f(\chi_2)-1) \HH(\chi_2) \mathcal{R}_2-5 s_b(s_2)+2) \, ,
\\
%%%%%%%%%%%%%%%%%%%%%
J^{\ndv \int \! \phi}_{20} \left( s_1 , s_2 , y ,\chi_2 \right) &=&
\frac{\Delta\chi_2^2 \HH_0^2 \Omega_{M0} D_1(\chi_2)     f_1 \HH_1 \mathcal{R}_1}{s_2 a(\chi_2)} (\chi_2 y-s_1)
\nonumber \\
&&
(s_2 (f(\chi_2)-1) \HH(\chi_2) \mathcal{R}_2-5 s_b(s_2)+2) \, ,
%%%%%%%%%%%%%%%%%%%%%
\\
J^{\ndv \int \! \phi}_{31} \left( s_1 , s_2 , y ,\chi_2 \right) &=&
-\frac{3 \chi_2^3 y f_0 \HH_0^3 \Omega_{M0} D_1(\chi_2)}{s_2 a(\chi_2)}
(\mathcal{R}_1-5 s_b(s_1)+2)
\nonumber \\
&&
(s_2 (f(\chi_2)-1) \HH(\chi_2) \mathcal{R}_2-5 s_b(s_2)+2) \, .
%%%%%%%%%%%%%%%%%%%%%
\eea

\subsubsection{Local Gravitational Potential x Integrated Gravitational Potential}
\be
\xi^{\phi \int \! \phi} \left( s_1 , s_2, \cos\theta \right) = D_1 \left( s_1 \right) \int_0^{s_2}\dd \chi_2
J^{\phi \int \! \phi}_{40} I^4_0\left( \Delta \chi_2 \right) 
+ \int_0^{s_2}\dd \chi_2  J^{\phi_0 \int \! \phi}_{40} I^4_0 \left(  \chi_2 \right)  \, ,
\ee
where
\bea
J^{\phi \int \! \phi}_{40} \left( s_1,s_2 , \chi_2 \right) &=&
\frac{3 \Delta\chi_2^4 \HH_0^2 \Omega_{M0} D_1(\chi_2) }{2 s_2 a(\chi_2) a(s_1)} (s_2 (f(\chi_2)-1) \HH(\chi_2) \mathcal{R}_2-5 s_b(s_2)+2) 
\nonumber \\
&&
\left(f_1 \left(2 a(s_1) (f_{\rm evo}(s_1)-3) \HH_1^2+3 \HH_0^2 \Omega_{M0}\right)+3 \HH_0^2 \Omega_{M0} (\mathcal{R}_1+5 s_b(s_1)-2)\right) \, ,
\nonumber \\
&&
\\
%%%%%%%%%%%%%%%%%
J^{\phi_0 \int \! \phi}_{40} \left( s_1,s_2 , \chi_2 \right) &=&
\frac{3 \chi_2^4 \HH_0^3 \Omega_{M0} D_1(\chi_2)}{2 s_1 s_2 a(\chi_2)} (s_2 (f(\chi_2)-1) \HH(\chi_2) \mathcal{R}_2-5 s_b(s_2)+2) 
\nonumber \\
&&
(\HH_0 s_1 (2 f_0-3 \Omega_{M0}) \mathcal{R}_1+2 f_0 (5 s_b(s_1)-2)) \, .
\eea

\section{The observer's velocity}
\label{sec:observer_vel_appendix}

The numerical results in section~\ref{sec:GR_effects_Pk} are computed in a frame determined by $v_{\rm obs } =0$. We now want to investigate the contribution of the observer velocity in more detail. In general, the correlation function will contain terms proportional to the variance of the velocity field
\be
\langle \ndv_0^2 \rangle = \frac{\cos\theta}{3} T_v^2 \left( 0 \right) \sigma_2 =  \frac{\cos\theta}{3} \HH_0^2 f_0^2 \sigma_2 
\ee
where $\cos \theta = \hat{\vs}_1 \cdot \hat{\vs}_2$, \ie they are pure dipoles.
From  definition of the Fourier space estimator, \eq{eq:def_estimator}, we get
\bea
\langle P^{\ndv_0^2}_0 \left( k \right) \rangle &=& \frac{1}{A} \int \frac{d\Omega_{\bk}}{4 \pi } \int \dd^3 s_1 \dd^3s_2  \frac{\cos\theta}{3} \HH_0^2 f_0^2 \sigma_2 \phi \left( \vs_1 \right) \phi \left( \vs_2 \right) e^{-i \bk \cdot \vs_2} e^{i \bk \cdot \vs_1}
\nonumber \\
&=&
\frac{1}{A} \int \dd^3 s_1 \dd^3 s_2  \frac{\cos\theta}{3} \HH_0^2 f_0^2 \sigma_2  \phi \left( \vs_1 \right) \phi \left( \vs_2 \right) \sum_\ell \left( 2\ell +1 \right) j_\ell \left( k s_1 \right) j_\ell \left( k s_2 \right) \mathcal{L}_\ell \left(\cos \theta \right) 
\nonumber \\
&&
\eea
Now considering a window function with an azimuthal symmetry, the above expression reduces to
\be
\label{eq:P_L_v0}
\langle P^{\ndv_0^2}_0 \left( k \right) \rangle = \frac{\HH^2_0 f_0^2 \sigma_2}{3 A} \sum_\ell \left( 2 \ell + 1\right) \phi_\ell^2 \left( k \right) g_\ell \left( \theta_{\rm max} \right) 
\ee
where 
\bea
\phi_\ell \left( k \right) &=& \int ds s^2 \phi\left( s \right) j_\ell \left( k s \right) 
\eea
and
\bea
g_\ell \left( \theta_{\rm \max} \right) &=& 2 \pi \int_0^{\theta_{\rm max}}\dd \theta_1 \sin \theta_1 \int_0^{\theta_{\rm max}}\dd \theta_2 \sin \theta_2 \int\dd \phi_1 \cos \theta \mathcal{L}_\ell \left( \cos \theta \right) \, .
\eea
For a full-sky survey the function, \ie $\theta_{\rm max}=\pi$, we simply have
\be
g_\ell\left( \pi \right) = \delta_{\ell 1} \frac{16 \pi^2}{3} \, .
\ee
and therefore
\bea
\langle P^{\ndv_0^2}_0 \left( k \right) \rangle_{\rm full-sky} 
=
\frac{\HH^2_0 f_0^2 \sigma_2}{3 A} \frac{16 \pi ^2}{k^6}
&&\left((k s_{\rm max} \sin (k s_{\rm max})+2 \cos (k s_{\rm max})
\right.
\nonumber \\
&& \left.
-k s_{\rm min} \sin (k s_{\rm min})-2 \cos (k s_{\rm min}) \right)^2 \, .
\eea
We can see that if $s_{\rm max} \gg s_{\rm min}$, the contribution to the power spectrum of the auto-correlation of the peculiar velocity evaluated at the observer position peaks at $k \sim \pi/s_{\max}$ and
\be
\langle P^{\ndv_0^2}_0 \left( \pi/s_{\rm max} \right) \rangle_{\rm full-sky} \sim \frac{\HH^2_0 f_0^2 \sigma_2}{3 A} \frac{256 s_{\rm max}^6}{\pi ^4} \, .
\ee
Analogously, for the case considered in the manuscript with $f_{\rm sky} =1/3$, we find
\be \label{eq:vel_max_powerspectrum}
\langle P^{\ndv_0^2}_0 \left( \pi/s_{\rm max} \right) \rangle
\sim \frac{\HH^2_0 f_0^2 \sigma_2}{3 A} \frac{s_{\rm max }^6}{2}  \, .
\ee
Therefore the contribution at the largest scales is given by the square of the Hankel transform of the window function together with some geometrical factor due the partial sky coverage.

\begin{figure}[!t]
\begin{center}
\includegraphics[width=1.09\textwidth]{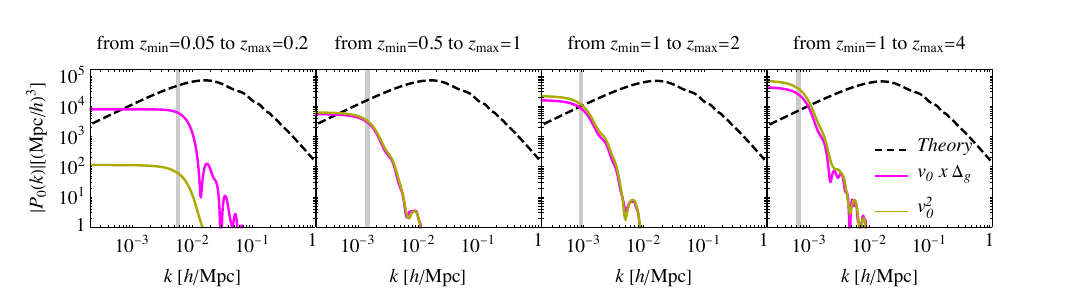}  
  \vspace{-0.75cm}
\caption{The contribution to the monopole of the power spectrum of the terms involving the observer velocity computed within linear theory.
\label{fig:vobs}}
\end{center}
\end{figure}

Interestingly, in full-sky this effect is purely sourced by the dipole of the correlation function. This is a spurious dipole induced by the observer motion inside a sphere defined by the full-sky window function in terms of real space coordinates. However, in a real survey the window function is determined in redshift space and therefore we need to account for its relation to the real space coordinates $\phi(z_1 ) \simeq \phi(s_1) + \frac{d\phi(s_1)}{ds_1} \frac{ds_1}{dz} \delta z_1$. This is the so-called Kaiser rocket effect~\cite{Kaiser:1987qv}, and it has been also generalized in a relativistic framework in Ref.~\cite{Bertacca:2019wyg}.
Fig.~\ref{fig:vobs} shows the effect of the observers' velocity on the large scale monopole of the power spectrum computed from \ref{eq:P_L_v0}. The shape of this contribution is completely determined by the angular and radial selection functions, as one can realize by noticing both auto- and cross- terms have the same shape.

So far we have implicitly assumed that the velocity field at the observer $v_{\rm obs}$ is a random variable drawn by the linear Gaussian distribution of cosmological perturbation statistics.
However, our motion with respect to the CMB frame can be measured from the CMB dipole~\cite{Kogut:1993ag,Lineweaver:1996xa,Hinshaw:2008kr,Akrami:2018vks} and from the modulation and aberration of the CMB anisotropies~\cite{Aghanim:2013suk,Akrami:2020nrk}. Our velocity with respect to the CMB is sourced by a combination of the linear gravitational field, usually identified with the Local Group Velocity, and short-scale non-linear effects, see e.g.~\cite{Nusser:2014sha}. 
In cross-correlating the observer velocity with the perturbations at the source position we need to account only for its linear component.

If the  observer velocity is known we can include its effect in our estimator. By assigning the measured value to $v_{\rm obs}$, we are in practise drawing a single cosmological realization~\cite{Mitsou:2019ocs,Desjacques:2020zue}. In this case we have
\bea
\langle P^{\ndv_0^2}_0 \left( k \right) \rangle &=& \frac{v_0^2}{A} \int \frac{d\Omega_{\bk}}{4 \pi } \int \dd^3 s_1 \dd^3s_2  
 \left( \hat \bv \cdot \hat \vs_1\right) 
 \left( \hat \bv \cdot \hat \vs_2\right)  \phi \left( \vs_1 \right) \phi \left( \vs_2 \right) e^{-i \bk \cdot \vs_2} e^{i \bk \cdot \vs_1}
 \nonumber \\
 &=& \frac{4 \pi}{A} v_0^2 \sum_{\ell m } \int ds_1 s_1^2 ds_2 s_2^2 \phi \left( s_1 \right) \phi \left( s_2 \right) j_\ell \left( k s_1 \right) j_\ell \left( k s_2 \right)  
 \nonumber 
\\
&&
\left\| \int\dd \Omega_{\vs_1} W \left( \hat \vs_1 \right)   \left( \hat \bv \cdot \hat \vs_1\right)  Y_{\ell m } \left( \hat \vs_1 \right)  \right\|^2 \, .
\eea
Now we compute
\bea
\int\dd \Omega_{\vs_1} W \left( \hat \vs_1 \right)   \left( \hat \bv \cdot \hat \vs_1\right)  Y_{\ell m } \left( \hat \vs_1 \right) 
&=& \frac{4 \pi}{3} \sum_{M=-1}^1 \int\dd \Omega_{\vs_1} W \left( \hat \vs_1 \right)   Y^*_{1M} \left( \hat \bv \right) Y_{1M} \left( \hat \vs_1 \right)   Y_{\ell m } \left( \hat \vs_1 \right) 
\nonumber \\
&=& \frac{4\pi }{3} \sum_{M=-1}^1 Y^*_{1M} \left( \hat \bv \right)   W_{\ell m M} \, ,
\eea
where
\be
W_{\ell m M} = \int\dd \Omega_{\vs_1} W \left( \hat \vs_1 \right)  Y_{1M} \left( \hat \vs_1 \right)   Y_{\ell m } \left( \hat \vs_1 \right) = \sum_{\ell' m'} W_{\ell' m'} \mathcal{G}_{1 \ell \ell'}^{M m m'}\, ,
\ee
from which it follows
\be
\left\| \int\dd \Omega_{\vs_1} W \left( \hat \vs_1 \right)   \left( \hat \bv \cdot \hat \vs_1\right)  Y_{\ell m } \left( \hat \vs_1 \right)  \right\|^2 = \frac{16 \pi^2}{9} 
\sum_{M=-1}^1\sum_{M'=-1}^1 Y^*_{1M} \left( \hat \bv \right) Y_{1M'} \left( \hat \bv \right) W_{\ell m M}W^*_{\ell m M'}  \, .
\ee
The final result reads as
\bea
\langle P^{\ndv_0^2}_0 \left( k \right) \rangle &=&
\frac{64 \pi^3}{9} \frac{v_0^2}{A} \sum_{\ell} \int ds_1 s_1^2 ds_2 s_2^2 \phi \left( s_1 \right) \phi \left( s_2 \right) j_\ell \left( k s_1 \right) j_\ell \left( k s_2 \right)  
\nonumber \\
&&
\sum_{M=-1}^1\sum_{M'=-1}^1 Y^*_{1M} \left( \hat \bv \right) Y_{1M'}\left( \hat \bv \right) \sum_m W_{\ell m M}W^*_{\ell m M'}
\nonumber \\
&=& \frac{64 \pi^3}{9} \frac{v_0^2}{A} \sum_{\ell} \phi^2_\ell \left( k \right) \sum_{M=-1}^1\sum_{M'=-1}^1 Y^*_{1M} \left( \hat \bv \right) Y_{1M'}\left( \hat \bv \right) \sum_m W_{\ell m M}W^*_{\ell m M'} \, . \quad 
\eea

\newpage
\bibliographystyle{JHEP}
\bibliography{}

\end{document}